\documentstyle[10pt,epsf,epsfig,dp_delphititle,oldlfont]{dp_delphi}
\makeindex
\def\DpPaperGroup{EP}
\def\DpPaperRef{2000-008}
\def\DpDate{14 January 2000}
\def\DpAuthors{DELPHI Collaboration}
\def\DpTitle{
{Search for charginos in ${\mathrm e}^+ {\mathrm e}^-$ 
 interactions at $\sqrt{s}=189$~GeV}}
\def\DpSubmit{(Phys. Lett. B479(2000)129)}
\begin{document}
%%%%%%%%%%%%%%%%%%%%%%%%%% They are a problem with Coll.Sty ?
\makeatletter
% Collapse citation numbers to ranges.  Non-numeric and undefined labels
% are handled.  No sorting is done.  E.g., 1,3,2,3,4,5,foo,1,2,3,?,4,5
% gives 1,3,2-5,foo,1-3,?,4,5
\newcount\@tempcntc
\def\@citex[#1]#2{\if@filesw\immediate\write\@auxout{\string\citation{#2}}\fi
  \@tempcnta\z@\@tempcntb\m@ne\def\@citea{}\@cite{\@for\@citeb:=#2\do
    {\@ifundefined
       {b@\@citeb}{\@citeo\@tempcntb\m@ne\@citea\def\@citea{,}{\bf ?}\@warning
       {Citation `\@citeb' on page \thepage \space undefined}}%
    {\setbox\z@\hbox{\global\@tempcntc0\csname b@\@citeb\endcsname\relax}%
     \ifnum\@tempcntc=\z@ \@citeo\@tempcntb\m@ne
       \@citea\def\@citea{,}\hbox{\csname b@\@citeb\endcsname}%
     \else
      \advance\@tempcntb\@ne
      \ifnum\@tempcntb=\@tempcntc
      \else\advance\@tempcntb\m@ne\@citeo
      \@tempcnta\@tempcntc\@tempcntb\@tempcntc\fi\fi}}\@citeo}{#1}}
\def\@citeo{\ifnum\@tempcnta>\@tempcntb\else\@citea\def\@citea{,}%
  \ifnum\@tempcnta=\@tempcntb\the\@tempcnta\else
   {\advance\@tempcnta\@ne\ifnum\@tempcnta=\@tempcntb \else \def\@citea{--}\fi
    \advance\@tempcnta\m@ne\the\@tempcnta\@citea\the\@tempcntb}\fi\fi}
 
\makeatother
%%%%%%%%%%%%%%%%%%%%%%%%%% ??????????????????????????????????
% Generate the title page
\begin{titlepage}
\pagenumbering{roman}
\CERNpreprint{\DpPaperGroup}{\DpPaperRef} % Reference of the paper
\date{{\small\DpDate}} % Date of the paper
\title{\DpTitle} % Title of the paper
\address{\DpAuthors} % General name of the author(s)
\begin{shortabs} % Start the abstract
\noindent
An update of the searches for  charginos and gravitinos is presented, based
on a data sample corresponding to  the 158 pb$^{-1}$ recorded by the DELPHI
detector in  1998, at a   centre-of-mass energy of  189 GeV. No  evidence for a
signal was  found. The lower mass limits are 4-5 GeV/$c^2$ higher than those
obtained at a centre-of-mass energy of 183 GeV.
The ($\mu$,M$_2$) MSSM domain excluded
by combining the chargino searches with neutralino searches at the ${\rm Z}$
resonance implies a  limit  on the mass of
the lightest neutralino which, for a heavy  sneutrino, is  constrained to be
above 31.0 GeV/$c^2$ for $\tan{\beta}\geq$ 1.
\end{shortabs}
\vfill
\begin{center}
\DpSubmit \ % Horrible hack to allow to have DpSubmit empty
\end{center}
\vfill
\clearpage
\headsep 10.0pt
\begingroup
% Commands to process the author names
%
\newcommand{\DpName}[2]{\hbox{#1$^{\ref{#2}}$},\hfill}
\newcommand{\DpNameTwo}[3]{\hbox{#1$^{\ref{#2},\ref{#3}}$},\hfill}
\newcommand{\DpNameThree}[4]{\hbox{#1$^{\ref{#2},\ref{#3},\ref{#4}}$},\hfill}
\newskip\Bigfill \Bigfill = 0pt plus 1000fill
\newcommand{\DpNameLast}[2]{\hbox{#1$^{\ref{#2}}$}\hspace{\Bigfill}}
%
%\small
\footnotesize
\noindent
\DpName{P.Abreu}{LIP}
\DpName{W.Adam}{VIENNA}
\DpName{T.Adye}{RAL}
\DpName{P.Adzic}{DEMOKRITOS}
\DpName{I.Ajinenko}{SERPUKHOV}
\DpName{Z.Albrecht}{KARLSRUHE}
\DpName{T.Alderweireld}{AIM}
\DpName{G.D.Alekseev}{JINR}
\DpName{R.Alemany}{VALENCIA}
\DpName{T.Allmendinger}{KARLSRUHE}
% ...
%\normalsize
%\endgroup
\titlefoot{Department of Physics and Astronomy, Iowa State
     University, Ames IA 50011-3160, USA
    \label{AMES}}
\titlefoot{Physics Department, Univ. Instelling Antwerpen,
     Universiteitsplein 1, B-2610 Antwerpen, Belgium \\
     \indent~~and IIHE, ULB-VUB,
     Pleinlaan 2, B-1050 Brussels, Belgium \\
     \indent~~and Facult\'e des Sciences,
     Univ. de l'Etat Mons, Av. Maistriau 19, B-7000 Mons, Belgium
    \label{AIM}}
\titlefoot{Physics Laboratory, University of Athens, Solonos Str.
     104, GR-10680 Athens, Greece
    \label{ATHENS}}
\titlefoot{Department of Physics, University of Bergen,
     All\'egaten 55, NO-5007 Bergen, Norway
    \label{BERGEN}}
\titlefoot{Dipartimento di Fisica, Universit\`a di Bologna and INFN,
     Via Irnerio 46, IT-40126 Bologna, Italy
    \label{BOLOGNA}}
\titlefoot{Centro Brasileiro de Pesquisas F\'{\i}sicas, rua Xavier Sigaud 150,
     BR-22290 Rio de Janeiro, Brazil \\
     \indent~~and Depto. de F\'{\i}sica, Pont. Univ. Cat\'olica,
     C.P. 38071 BR-22453 Rio de Janeiro, Brazil \\
     \indent~~and Inst. de F\'{\i}sica, Univ. Estadual do Rio de Janeiro,
     rua S\~{a}o Francisco Xavier 524, Rio de Janeiro, Brazil
    \label{BRASIL}}
\titlefoot{Comenius University, Faculty of Mathematics and Physics,
     Mlynska Dolina, SK-84215 Bratislava, Slovakia
    \label{BRATISLAVA}}
\titlefoot{Coll\`ege de France, Lab. de Physique Corpusculaire, IN2P3-CNRS,
     FR-75231 Paris Cedex 05, France
    \label{CDF}}
\titlefoot{CERN, CH-1211 Geneva 23, Switzerland
    \label{CERN}}
\titlefoot{Institut de Recherches Subatomiques, IN2P3 - CNRS/ULP - BP20,
     FR-67037 Strasbourg Cedex, France
    \label{CRN}}
\titlefoot{Now at DESY-Zeuthen, Platanenallee 6, D-15735 Zeuthen, Germany
    \label{DESY}}
\titlefoot{Institute of Nuclear Physics, N.C.S.R. Demokritos,
     P.O. Box 60228, GR-15310 Athens, Greece
    \label{DEMOKRITOS}}
\titlefoot{FZU, Inst. of Phys. of the C.A.S. High Energy Physics Division,
     Na Slovance 2, CZ-180 40, Praha 8, Czech Republic
    \label{FZU}}
\titlefoot{Dipartimento di Fisica, Universit\`a di Genova and INFN,
     Via Dodecaneso 33, IT-16146 Genova, Italy
    \label{GENOVA}}
\titlefoot{Institut des Sciences Nucl\'eaires, IN2P3-CNRS, Universit\'e
     de Grenoble 1, FR-38026 Grenoble Cedex, France
    \label{GRENOBLE}}
\titlefoot{Helsinki Institute of Physics, HIP,
     P.O. Box 9, FI-00014 Helsinki, Finland
    \label{HELSINKI}}
\titlefoot{Joint Institute for Nuclear Research, Dubna, Head Post
     Office, P.O. Box 79, RU-101 000 Moscow, Russian Federation
    \label{JINR}}
\titlefoot{Institut f\"ur Experimentelle Kernphysik,
     Universit\"at Karlsruhe, Postfach 6980, DE-76128 Karlsruhe,
     Germany
    \label{KARLSRUHE}}
\titlefoot{Institute of Nuclear Physics and University of Mining and Metalurgy,
     Ul. Kawiory 26a, PL-30055 Krakow, Poland
    \label{KRAKOW}}
\titlefoot{Universit\'e de Paris-Sud, Lab. de l'Acc\'el\'erateur
     Lin\'eaire, IN2P3-CNRS, B\^{a}t. 200, FR-91405 Orsay Cedex, France
    \label{LAL}}
\titlefoot{School of Physics and Chemistry, University of Lancaster,
     Lancaster LA1 4YB, UK
    \label{LANCASTER}}
\titlefoot{LIP, IST, FCUL - Av. Elias Garcia, 14-$1^{o}$,
     PT-1000 Lisboa Codex, Portugal
    \label{LIP}}
\titlefoot{Department of Physics, University of Liverpool, P.O.
     Box 147, Liverpool L69 3BX, UK
    \label{LIVERPOOL}}
\titlefoot{LPNHE, IN2P3-CNRS, Univ.~Paris VI et VII, Tour 33 (RdC),
     4 place Jussieu, FR-75252 Paris Cedex 05, France
    \label{LPNHE}}
\titlefoot{Department of Physics, University of Lund,
     S\"olvegatan 14, SE-223 63 Lund, Sweden
    \label{LUND}}
\titlefoot{Universit\'e Claude Bernard de Lyon, IPNL, IN2P3-CNRS,
     FR-69622 Villeurbanne Cedex, France
    \label{LYON}}
\titlefoot{Univ. d'Aix - Marseille II - CPP, IN2P3-CNRS,
     FR-13288 Marseille Cedex 09, France
    \label{MARSEILLE}}
\titlefoot{Dipartimento di Fisica, Universit\`a di Milano and INFN-MILANO,
     Via Celoria 16, IT-20133 Milan, Italy
    \label{MILANO}}
\titlefoot{Dipartimento di Fisica, Univ. di Milano-Bicocca and
     INFN-MILANO, Piazza delle Scienze 2, IT-20126 Milan, Italy
    \label{MILANO2}}
\titlefoot{Niels Bohr Institute, Blegdamsvej 17,
     DK-2100 Copenhagen {\O}, Denmark
    \label{NBI}}
\titlefoot{IPNP of MFF, Charles Univ., Areal MFF,
     V Holesovickach 2, CZ-180 00, Praha 8, Czech Republic
    \label{NC}}
\titlefoot{NIKHEF, Postbus 41882, NL-1009 DB
     Amsterdam, The Netherlands
    \label{NIKHEF}}
\titlefoot{National Technical University, Physics Department,
     Zografou Campus, GR-15773 Athens, Greece
    \label{NTU-ATHENS}}
\titlefoot{Physics Department, University of Oslo, Blindern,
     NO-1000 Oslo 3, Norway
    \label{OSLO}}
\titlefoot{Dpto. Fisica, Univ. Oviedo, Avda. Calvo Sotelo
     s/n, ES-33007 Oviedo, Spain
    \label{OVIEDO}}
\titlefoot{Department of Physics, University of Oxford,
     Keble Road, Oxford OX1 3RH, UK
    \label{OXFORD}}
\titlefoot{Dipartimento di Fisica, Universit\`a di Padova and
     INFN, Via Marzolo 8, IT-35131 Padua, Italy
    \label{PADOVA}}
\titlefoot{Rutherford Appleton Laboratory, Chilton, Didcot
     OX11 OQX, UK
    \label{RAL}}
\titlefoot{Dipartimento di Fisica, Universit\`a di Roma II and
     INFN, Tor Vergata, IT-00173 Rome, Italy
    \label{ROMA2}}
\titlefoot{Dipartimento di Fisica, Universit\`a di Roma III and
     INFN, Via della Vasca Navale 84, IT-00146 Rome, Italy
    \label{ROMA3}}
\titlefoot{DAPNIA/Service de Physique des Particules,
     CEA-Saclay, FR-91191 Gif-sur-Yvette Cedex, France
    \label{SACLAY}}
\titlefoot{Instituto de Fisica de Cantabria (CSIC-UC), Avda.
     los Castros s/n, ES-39006 Santander, Spain
    \label{SANTANDER}}
\titlefoot{Dipartimento di Fisica, Universit\`a degli Studi di Roma
     La Sapienza, Piazzale Aldo Moro 2, IT-00185 Rome, Italy
    \label{SAPIENZA}}
\titlefoot{Inst. for High Energy Physics, Serpukov
     P.O. Box 35, Protvino, (Moscow Region), Russian Federation
    \label{SERPUKHOV}}
\titlefoot{J. Stefan Institute, Jamova 39, SI-1000 Ljubljana, Slovenia
     and Laboratory for Astroparticle Physics,\\
     \indent~~Nova Gorica Polytechnic, Kostanjeviska 16a, SI-5000 Nova Gorica, Slovenia, \\
     \indent~~and Department of Physics, University of Ljubljana,
     SI-1000 Ljubljana, Slovenia
    \label{SLOVENIJA}}
\titlefoot{Fysikum, Stockholm University,
     Box 6730, SE-113 85 Stockholm, Sweden
    \label{STOCKHOLM}}
\titlefoot{Dipartimento di Fisica Sperimentale, Universit\`a di
     Torino and INFN, Via P. Giuria 1, IT-10125 Turin, Italy
    \label{TORINO}}
\titlefoot{Dipartimento di Fisica, Universit\`a di Trieste and
     INFN, Via A. Valerio 2, IT-34127 Trieste, Italy \\
     \indent~~and Istituto di Fisica, Universit\`a di Udine,
     IT-33100 Udine, Italy
    \label{TU}}
\titlefoot{Univ. Federal do Rio de Janeiro, C.P. 68528
     Cidade Univ., Ilha do Fund\~ao
     BR-21945-970 Rio de Janeiro, Brazil
    \label{UFRJ}}
\titlefoot{Department of Radiation Sciences, University of
     Uppsala, P.O. Box 535, SE-751 21 Uppsala, Sweden
    \label{UPPSALA}}
\titlefoot{IFIC, Valencia-CSIC, and D.F.A.M.N., U. de Valencia,
     Avda. Dr. Moliner 50, ES-46100 Burjassot (Valencia), Spain
    \label{VALENCIA}}
\titlefoot{Institut f\"ur Hochenergiephysik, \"Osterr. Akad.
     d. Wissensch., Nikolsdorfergasse 18, AT-1050 Vienna, Austria
    \label{VIENNA}}
\titlefoot{Inst. Nuclear Studies and University of Warsaw, Ul.
     Hoza 69, PL-00681 Warsaw, Poland
    \label{WARSZAWA}}
\titlefoot{Fachbereich Physik, University of Wuppertal, Postfach
     100 127, DE-42097 Wuppertal, Germany
    \label{WUPPERTAL}}
\endgroup
\clearpage
\end{titlepage}
\headsep 30.0pt
%%%%%%%%%%%%%%%%%%%%%%%%%
%
% Change for the document body
%\pagestyle{heading} % for page numbering
\pagenumbering{arabic} % page numbering in number
\renewcommand{\thefootnote}{\fnsymbol{footnote}} % symbolic footnote marks
\setcounter{footnote}{1} %
\large
% document.tex
%unch
\def\leqsim{\mathbin{\;\raise1pt\hbox{$<$}\kern-8pt\lower3pt\hbox{\small$\sim$}\;}}
\def\geqsim{\mathbin{\;\raise1pt\hbox{$>$}\kern-8pt\lower3pt\hbox{\small$\sim$}\;}}
\newcommand{\dfrac}[2]{\frac{\displaystyle #1}{\displaystyle #2}}
\renewcommand\topfraction{1.}
\renewcommand\bottomfraction{1.}
\renewcommand\floatpagefraction{0.}
\renewcommand\textfraction{0.}
% Charginos and Neutralinos :
\def\MXN#1{\mbox{$ M_{\tilde{\chi}^0_#1}                                $}}
\def\MXNN#1#2{\mbox{$ M_{\tilde{\chi}^0_{#1,#2}}                        $}}
\def\MXNNN#1#2#3{\mbox{$ M_{\tilde{\chi}^0_{#1,#2,#3}}                  $}}
\def\MXC#1{\mbox{$ M_{\tilde{\chi}^\pm _#1}                            $}}
\def\XP#1{\mbox{$ \tilde{\chi}^+_#1                                     $}}
\def\XM#1{\mbox{$ \tilde{\chi}^-_#1                                     $}}
\def\XPM#1{\mbox{$ \tilde{\chi}^\pm _#1                                $}}
\def\XN#1{\mbox{$ \tilde{\chi}^0_#1                                     $}}
\def\XNN#1#2{\mbox{$ \tilde{\chi}^0_{#1,#2}                             $}}
\def\XNNN#1#2#3{\mbox{$ \tilde{\chi}^0_{#1,#2,#3}                       $}}
\def\MXn{\mbox{$ M_{\tilde{\chi}^0}                                     $}}
\def\MXc{\mbox{$ M_{\tilde{\chi}^{\pm}}                                 $}}
\def\p#1{\mbox{$ \mbox{\bf p}_1                                         $}}
\newcommand{\Gino}    {\mbox{$ \tilde{\mathrm G}                           $}}
\newcommand{\tanb}    {\mbox{$ \tan \beta                                  $}}
\newcommand{\smu}     {\mbox{$ \tilde\mu                                  $}}
\newcommand{\msmu}    {\mbox{$ M_{\tilde\mu }                             $}}
\newcommand{\msmur}   {\mbox{$ M_{\tilde\mu _R}                           $}}
\newcommand{\msmul}   {\mbox{$ M_{\tilde\mu _L}                           $}}
\newcommand{\sel}     {\mbox{$ \tilde{\mathrm e}                           $}}
\newcommand{\msel}    {\mbox{$ M_{\tilde{\mathrm e}}                       $}}
\newcommand{\snu}     {\mbox{$ \tilde\nu                                   $}}
\newcommand{\msnu}    {\mbox{$ M_{\tilde\nu}                               $}}
\newcommand{\msell}   {\mbox{$ M_{\tilde{\mathrm e}_L}                     $}}
\newcommand{\mselr}   {\mbox{$ M_{\tilde{\mathrm e}_R}                     $}}
\newcommand{\sfe}     {\mbox{$ \tilde{\mathrm f}                           $}}
\newcommand{\msfe}    {\mbox{$ M_{\tilde{\mathrm f}}                       $}}
\newcommand{\sle}     {\mbox{$ \tilde{\ell}                                $}}
\newcommand{\msle}    {\mbox{$ M_{\tilde{\ell}}                            $}}
\newcommand{\stq}     {\mbox{$ \tilde {\mathrm t}                          $}}
\newcommand{\mstq}    {\mbox{$ M_{\tilde {\mathrm t}}                      $}}
\newcommand{\sbq}     {\mbox{$ \tilde {\mathrm b}                          $}}
\newcommand{\msbq}    {\mbox{$ M_{\tilde {\mathrm b}}                      $}}
\newcommand{\An}      {\mbox{$ {\mathrm A}^0                               $}}
\newcommand{\hn}      {\mbox{$ {\mathrm h}^0                               $}}
\newcommand{\Zn}      {\mbox{$ {\mathrm Z}                                 $}}
\newcommand{\Zstar}   {\mbox{$ {\mathrm Z}^*                               $}}
\newcommand{\Hn}      {\mbox{$ {\mathrm H}^0                               $}}
\newcommand{\HP}      {\mbox{$ {\mathrm H}^+                               $}}
\newcommand{\HM}      {\mbox{$ {\mathrm H}^-                               $}}
\newcommand{\Wp}      {\mbox{$ {\mathrm W}^+                               $}}
\newcommand{\Wm}      {\mbox{$ {\mathrm W}^-                               $}}
\newcommand{\Wstar}   {\mbox{$ {\mathrm W}^*                               $}}
\newcommand{\WW}      {\mbox{$ {\mathrm W}^+{\mathrm W}^-                  $}}
\newcommand{\ZZ}      {\mbox{$ {\mathrm Z}{\mathrm Z}                      $}}
\newcommand{\HZ}      {\mbox{$ {\mathrm H}^0 {\mathrm Z}                   $}}
\newcommand{\GW}      {\mbox{$ \Gamma_{\mathrm W}                          $}}
\newcommand{\Zg}      {\mbox{$ \Zn \gamma                                  $}}
\newcommand{\sqs}     {\mbox{$ \sqrt{s}                                    $}}
\newcommand{\epm}     {\mbox{$ {\mathrm e}^\pm                            $}}
\newcommand{\ee}      {\mbox{$ {\mathrm e}^+ {\mathrm e}^-                 $}}
\newcommand{\mumu}    {\mbox{$ \mu ^+ \mu ^-                                 $}}
\newcommand{\eeto}    {\mbox{$ {\mathrm e}^+ {\mathrm e}^-\! \to\          $}}
\newcommand{\ellell}  {\mbox{$ \ell^+ \ell^-                               $}}
\newcommand{\eeWW}    {\mbox{$ \ee \rightarrow \WW                         $}}
\newcommand{\eV}      {\mbox{$ {\mathrm{eV}}                               $}}
\newcommand{\eVc}     {\mbox{$ {\mathrm{eV}}/c                             $}}
\newcommand{\eVcc}    {\mbox{$ {\mathrm{eV}}/c^2                           $}}
\newcommand{\MeV}     {\mbox{$ {\mathrm{MeV}}                              $}}
\newcommand{\MeVc}    {\mbox{$ {\mathrm{MeV}}/c                            $}}
\newcommand{\MeVcc}   {\mbox{$ {\mathrm{MeV}}/c^2                          $}}
\newcommand{\GeV}     {\mbox{$ {\mathrm{GeV}}                              $}}
\newcommand{\GeVc}    {\mbox{$ {\mathrm{GeV}}/c                            $}}
\newcommand{\GeVcc}   {\mbox{$ {\mathrm{GeV}}/c^2                          $}}
\newcommand{\TeV}     {\mbox{$ {\mathrm{TeV}}                              $}}
\newcommand{\TeVc}    {\mbox{$ {\mathrm{TeV}}/c                            $}}
\newcommand{\TeVcc}   {\mbox{$ {\mathrm{TeV}}/c^2                          $}}
\newcommand{\pbi}     {\mbox{$ {\mathrm{pb}}^{-1}                          $}}
\newcommand{\MZ}      {\mbox{$ M_{\mathrm Z}                               $}}
\newcommand{\MW}      {\mbox{$ M_{\mathrm W}                               $}}
\newcommand{\MA}      {\mbox{$ m_{\mathrm A}                               $}}
\newcommand{\GF}      {\mbox{$ {\mathrm G}_{\mathrm F}                     $}}
\newcommand{\MH}      {\mbox{$ m_{{\mathrm H}^0}                           $}}
\newcommand{\MHP}     {\mbox{$ m_{{\mathrm H}^\pm }                         $}}
\newcommand{\MSH}     {\mbox{$ m_{{\mathrm h}^0}                           $}}
\newcommand{\MT}      {\mbox{$ m_{\mathrm t}                               $}}
\newcommand{\GZ}      {\mbox{$ \Gamma_{{\mathrm Z} }                       $}}
\newcommand{\SS}      {\mbox{$ \mathrm S                                   $}}
\newcommand{\TT}      {\mbox{$ \mathrm T                                   $}}
\newcommand{\UU}      {\mbox{$ \mathrm U                                   $}}
\newcommand{\alphmz}  {\mbox{$ \alpha (m_{{\mathrm Z}})                    $}}
\newcommand{\alphas}  {\mbox{$ \alpha_{\mathrm s}                          $}}
\newcommand{\alphmsb} {\mbox{$ \alphas (m_{\mathrm Z})
                               _{\overline{\mathrm{MS}}}                   $}}
\newcommand{\alphbar} {\mbox{$ \overline{\alpha}_{\mathrm s}               $}}
\newcommand{\Ptau}    {\mbox{$ P_{\tau}                                    $}}
\newcommand{\mean}[1] {\mbox{$ \left\langle #1 \right\rangle               $}}
\newcommand{\dgree}   {\mbox{$ ^\circ                                      $}}
\newcommand{\qqg}     {\mbox{$ {\mathrm q}\bar{\mathrm q}\gamma            $}}
\newcommand{\Wev}     {\mbox{$ {\mathrm{W e}} \nu_{\mathrm e}              $}}
\newcommand{\Zvv}     {\mbox{$ \Zn \nu \bar{\nu}                           $}}
\newcommand{\Zee}     {\mbox{$ \Zn \ee                                     $}}
\newcommand{\ctw}     {\mbox{$ \cos\theta_{\mathrm W}                      $}}
\newcommand{\thw}     {\mbox{$ \theta_{\mathrm W}                          $}}
\newcommand{\thetabar}{\mbox{$ \theta^*                                    $}}
\newcommand{\phibar}  {\mbox{$ \phi^*                                      $}}
\newcommand{\thetapl} {\mbox{$ \theta_+                                    $}}
\newcommand{\phipl}   {\mbox{$ \phi_+                                      $}}
\newcommand{\thetamin}{\mbox{$ \theta_-                                    $}}
\newcommand{\phimin}  {\mbox{$ \phi_-                                      $}}
\newcommand{\ds}      {\mbox{$ {\mathrm d} \sigma                          $}}
\def    \ll           {\mbox{$\ell \ell                                    $}}
\def    \jjl          {\mbox{$jj \ell                                      $}}
\def    \jj           {\mbox{$jj                                           $}}
\def   \jjjj          {\mbox{${\it jets}                                   $}}
\def   \rad           {\mbox{${\it rad}                                    $}}
\def    \llt          {\mbox{$\ell \ell_t                                  $}}
\def    \jjlt         {\mbox{$jj \ell_t                                    $}}
\def    \jjt          {\mbox{$jj_t                                         $}}
\def   \jjjjt         {\mbox{${\it jets_t}                                 $}}
\def   \radt          {\mbox{${\it rad_t}                                  $}}
\newcommand{\jjlv}    {\mbox{$ j j \ell \nu                                $}}
\newcommand{\jjvv}    {\mbox{$ j j \nu \bar{\nu}                           $}}
\newcommand{\qqvv}    {\mbox{$ \mathrm{q \bar{q}} \nu \bar{\nu}            $}}
\newcommand{\qqll}    {\mbox{$ \mathrm{q \bar{q}} \ell \bar{\ell}          $}}
\newcommand{\jjll}    {\mbox{$ j j \ell \bar{\ell}                         $}}
\newcommand{\lvlv}    {\mbox{$ \ell \nu \ell \nu                           $}}
\newcommand{\dz}      {\mbox{$ \delta g_{\mathrm{W W Z}    }               $}}
\newcommand{\pT}      {\mbox{$ p_{\mathrm{T}}                              $}}
\newcommand{\ptr}     {\mbox{$ p_{\perp}                                   $}}
\newcommand{\ptrjet}  {\mbox{$ p_{\perp {\mathrm{jet}}}                    $}}
\newcommand{\Wvis}    {\mbox{$ {\mathrm W}_{\mathrm{vis}}                  $}}
\newcommand{\gamgam}  {\mbox{$ \gamma \gamma                               $}}
\newcommand{\qaqb}    {\mbox{$ {\mathrm q}_1 \bar{\mathrm q}_2             $}}
\newcommand{\qcqd}    {\mbox{$ {\mathrm q}_3 \bar{\mathrm q}_4             $}}
\newcommand{\bbbar}   {\mbox{$ {\mathrm b}\bar{\mathrm b}                  $}}
\newcommand{\ffbar}   {\mbox{$ {\mathrm f}\bar{\mathrm f}                  $}}
\newcommand{\ffbars}   {\mbox{$ {\mathrm f}_2\bar{\mathrm f}_2             $}}

\newcommand{\ffbarp}  {\mbox{$ {\mathrm f}\bar{\mathrm f}'                 $}}
\newcommand{\ffbarpf}  {\mbox{$ {\mathrm f}_1\bar{\mathrm f}_1'            $}}

\newcommand{\qqbar}   {\mbox{$ {\mathrm q}\bar{\mathrm q}                  $}}
\newcommand{\nunubar} {\mbox{$ {\nu}\bar{\nu}                              $}}
\newcommand{\qqbarp}  {\mbox{$ {\mathrm q'}\bar{\mathrm q}'                $}}
\newcommand{\djoin}   {\mbox{$ d_{\mathrm{join}}                           $}}
\newcommand{\mErad}   {\mbox{$ \left\langle E_{\mathrm{rad}} \right\rangle $}}
%%%%%%%%%%%%%%%%%%%%%%%
% End of Declarations S.K %
%%%%%%%%%%%%%%%%%%%%%%%
%\newcommand{\bibit}{\nineit}
%\newcommand{\bibbf}{\ninebf}
\newcommand{\Lum}{${\cal L}\;$}
\newcommand{\lum}{{\cal L}}
\newcommand{\Cms}{$\mbox{ cm}^{-2} \mbox{ s}^{-1}\;$}
\newcommand{\cms}{\mbox{ cm}^{-2} \mbox{ s}^{-1}\;}
\newcommand{\Ecms}    {\mbox{$ E_{\mathrm{\small cms}}                      $}}
\newcommand{\Evis}    {\mbox{$ E_{\mathrm{\small vis}}                      $}}
\newcommand{\Erad}    {\mbox{$ E_{\mathrm{\small rad}}                      $}}
\newcommand{\Mvis}    {\mbox{$ M_{\mathrm{\small vis}}                      $}}
\newcommand{\pvis}    {\mbox{$ p_{\mathrm{\small vis}}                      $}}
\newcommand{\Minv}    {\mbox{$ M_{\mathrm{\small inv}}                      $}}
\newcommand{\pmiss}   {\mbox{$ p_{\mathrm{\small miss}}                     $}}
\newcommand{\ptmiss}  {\mbox{$ p_T^{\mathrm{\small miss}}                   $}}
\newcommand{\ptpair}  {\mbox{$ p_T^{\mathrm{\small pair}}                   $}}
\newcommand{\Mhfit}{\; \hat{m}_{H^0} }
\newcommand{\bl}      {\mbox{\ \ \ \ \ \ \ \ \ \ } }
%%%%%%%%%%%%%%%%%%%%%%%
% End of Declarations J.M %
%%%%%%%%%%%%%%%%%%%%%%%
\newcommand{\Zto}   {\mbox{$\mathrm Z^0 \to$}}
\newcommand{\etal}  {\mbox{\it et al.}}
\def\NPB#1#2#3{{\rm Nucl.~Phys.} {\bf{B#1}} (19#2) #3}
\def\PLB#1#2#3{{\rm Phys.~Lett.} {\bf{B#1}} (#2) #3}
\def\PRD#1#2#3{{\rm Phys.~Rev.} {\bf{D#1}} (19#2) #3}
\def\PRL#1#2#3{{\rm Phys.~Rev.~Lett.} {\bf{#1}} (19#2) #3}
\def\ZPC#1#2#3{{\rm Z.~Phys.} {\bf C#1} (19#2) #3}
\def\PTP#1#2#3{{\rm Prog.~Theor.~Phys.} {\bf#1}  (19#2) #3}
\def\MPL#1#2#3{{\rm Mod.~Phys.~Lett.} {\bf#1} (19#2) #3}
\def\PR#1#2#3{{\rm Phys.~Rep.} {\bf#1} (19#2) #3}
\def\RMP#1#2#3{{\rm Rev.~Mod.~Phys.} {\bf#1} (19#2) #3}
\def\HPA#1#2#3{{\rm Helv.~Phys.~Acta} {\bf#1} (19#2) #3}
\def\NIMA#1#2#3{{\rm Nucl.~Instr.~and~Meth.} {\bf#1} (19#2) #3}
\def\CPC#1#2#3{{\rm Comp.~Phys.~Comm.} {\bf#1} (19#2) #3}
% Imported from chargino paper
\def    \DM          {\mbox{$\Delta M$}}
\def    \missEt      {\ifmmode{/\mkern-11mu E_t}\else{${/\mkern-11mu E_t}$}\fi}
\def    \missE       {\ifmmode{/\mkern-11mu E}\else{${/\mkern-11mu E}$}\fi}
\def    \missp       {\ifmmode{/\mkern-11mu p}\else{${/\mkern-11mu p}$}\fi}
\def    \misspt      {\ifmmode{/\mkern-11mu p_t}\else{${/\mkern-11mu p_t}$}\fi}
\def    \DML         {\mbox{5~GeV $<\Delta M<$ 10~GeV}}
\def    \rs          {\mbox{$\sqrt{s}$}}
\def    \msneu       {\mbox{$m_{\tilde{\nu}}$}}

\section{Introduction \label{sec:INTRO}}

  In 1998, the LEP centre-of-mass energy reached 189~GeV, and
the DELPHI experiment collected an integrated luminosity of 158~\pbi.
These data have been analysed to search for charginos, supersymmetric partners
of Higgs and gauge bosons, 
predicted by supersymmetric (SUSY) models \cite{SUSY2}.

 A description of the parts of the DELPHI detector
relevant to the present paper can be found in \cite{PAP172}, while
a complete description is given in ~\cite{DELPHI}.

 The results obtained at the same centre-of-mass energy by other LEP collaborations, 
on similar searches, are described in ~\cite{OTHC189}.

  The conservation of R-parity, implying a stable lightest
supersymmetric particle (LSP), is assumed. 
This also means that charginos are pair produced in \ee\ collisions.
The analysis was performed in the framework of the Minimal Supersymmetric extension of the
Standard Model (MSSM), with
universal parameters at the high mass scale typical of Grand Unified Theories
(GUT's) \cite{SUSY2}.
The parameters of this model relevant to the present searches are the masses
$M_1$ and $M_2$ of the gaugino sector (which are assumed to satisfy the GUT
relation $M_1 = \frac{5}{3}\tan^2\theta_W\! M_2 \approx 0.5 M_2 $ at the
electroweak scale), the universal mass $m_0$ of the scalar fermion sector,
the Higgs mass parameter $\mu $, and the ratio \tanb\ of the vacuum expectation
values of the two Higgs doublets. In this paper it is assumed that $m_0=$1 TeV. 
Scalar mass unification is assumed,
except for the sneutrino mass which is considered to be a free parameter. 
As in Ref.~\cite{PAP172} both cases where either the lightest neutralino (\XN{1}) 
or the gravitino (\Gino) is the LSP are considered.

   In the former case, the decay of the charginos is \XPM{1}$\rightarrow$\XN{1}\ffbarp\ (\ffbarp\ can be quarks or leptons) 
and the events are characterised by missing energy carried by the escaping \XN{1}\ . 
In some areas of the parameter space, the charginos can decay to heavier neutralinos giving rise to a cascade effect:
\XPM{1}$\rightarrow$\XN{2}\ffbarpf$\rightarrow$\XN{1} \ffbarpf\ \ffbars\ (\ffbarpf\ \ffbars\ can be quarks or leptons).
The decay  \XN{2}$\rightarrow$\XN{1}$\gamma$ may occur for small $\mu \approx -M_2$.
So the following decay channels were defined:
\begin{itemize}
\item{The $leptonic$\ channel (\ll ): the decay products are only leptons and the LSPs.}
\item{The $hadronic$\ channel (\jjjj): the decay products are only quarks and the LSPs.}
\item{The $semi$-$leptonic$\ channel (\jjl): the decay products are quarks, leptons and the LSPs.}
\item{The $radiative$\ channel (\rad): there is at least one isolated photon among the decay products.}
\end{itemize}
In this scenario the likelihood ratio method \cite{ANDER} was used to optimize 
the search for charginos.
An overview of this method and details of the implementation are given in~\ref{sub:LIKELI}.

  If the gravitino is the LSP, the decay $\XN{1} \rightarrow \Gino \gamma$ is
possible \cite{OLDGRAV,GRAVITINO,AMBROSANIO}. If the gravitino is
sufficiently light (with a mass below about 10~eV/$c^2$ \cite{AMBROSANIO}), 
this decay takes place within
the detector. As gravitinos escape detection, the typical signature
of these SUSY events is missing energy and isolated photons.
The selection criteria already used at a centre-of-mass energy of 183 \GeV\ are 
applied in this scenario.
The detailed description of the analysis can be found in Ref.~\cite{PAP172}.

\section{Event generators\label{sec:SAMPLES}}

  To evaluate the signal efficiencies and background contaminations, events were
generated using several different programs. All relied on {\tt JETSET}
7.4~\cite{JETSET}, tuned to LEP~1 data ~\cite{TUNE}, for quark fragmentation.

  The program {\tt SUSYGEN}~\cite{SUSYGEN} was used to generate events with chargino production and
decay in both the neutralino LSP and the gravitino LSP scenarios,
and to calculate masses, cross-sections and branching ratios for each adopted parameter set.
 These agree with the calculations of Ref.~\cite{AM}. Details of the signal samples generated
are given in section~\ref{sec:RES}.

  The background process \eeto\qqbar ($n\gamma$) was generated with
{\tt PYTHIA 5.7} \cite{JETSET}, while {\tt DYMU3}~\cite{DYMU3} and
{\tt KORALZ 4.2}~\cite{KORALZ} were used
for $\mu ^+\mu ^-(\gamma)$ and $\tau^+\tau^-(\gamma)$,
respectively. The generator of Ref.~\cite{BAFO} was used for \eeto\ee\ events.
Processes leading to four-fermion final states,
$(\Zn/\gamma)^*(\Zn/\gamma)^*$, $\Wp \Wm $, \Wev\ and \Zee,
were generated using {\tt EXCALIBUR}~\cite{EXCALIBUR} and {\tt GRC4F}~\cite{GRACE}. 
 
  Two-photon interactions leading to hadronic final states were generated using
{\tt TWOGAM}~\cite{TWOGAM}, separating the VDM (Vector Dominance Model) and QCD components.
The generators of \mbox{Berends}, Daverveldt and Kleiss~\cite{BDK} were used for 
the QPM (Quark Parton Model) component and for leptonic final states.

  The generated signal and background events were passed through the
detailed simulation of the DELPHI detector~\cite{DELPHI} and then processed
with the same reconstruction and analysis programs as real data events.
The number of simulated events from different background processes was
several times (a factor varying from 2 to 140 depending on the background
process) the number of real events recorded.

\section{Event selections \label{sec:EVSEL}}

  The criteria used to select events were defined on the
basis of the simulated signal and background events. The selections 
for charged and neutral particles were similar to those presented in 
\cite{PAP172}, requiring charged particles to
have momentum above 100~\MeVc\ and to extrapolate back to
within 5~cm of the main vertex in the transverse plane, and to within 
twice this distance in the longitudinal direction.
Calorimeter energy clusters above 100~\MeV\ were taken as
neutral particles if not associated to any charged particle track.
The particle selection was followed by different event
selections for the different signal topologies considered in the application
of the likelihood ratio method, which was used in the stable \XN{1}\ case. 
The detailed description of the analysis done in the unstable \XN{1}\ case
can be found in Ref.~\cite{PAP172}.

\subsection{The likelihood ratio method \label{sub:LIKELI}}

In the likelihood ratio method used, several discriminating variables
are combined into one on the basis of their one-dimensional probability
density functions (pdf). If the variables used are independent, this gives
the best possible background suppression for a given signal efficiency~\cite{ANDER}.
%KH The likelihood ratio method consists of combining several
%KH discriminating variables into one. This method, if the variables used are independent, 
%KH gives the best possible background suppression for a given signal 
%KH efficiency. The proof of this statement can be found in \cite{ANDER}.
%KH 
For a set of variables $\left\{ x_{i}\right\} $, the pdfs
of these variables are estimated by normalised 
frequency distributions for the signal and the background samples.
We denote the pdfs of these variables $f_{i}^{S}(x_{i})$ for
the signal events and $f_{i}^{B}(x_{i})$ for the background events submitted 
to the same selection criteria.
The likelihood ratio function is defined as $\mathcal{L}_{R}$ $ = \prod\limits_{i=1}^{n}\frac{%
f_{i}^{S}(x_{i})}{f_{i}^{B}(x_{i})}$. Events with $\mathcal{L}_{R}>\mathcal{L}_{R_{CUT}}$\ are selected
as candidate signal events.
  The optimal set of variables and the value of $\mathcal{L}_{R_{CUT}}$ were defined 
in order to minimise the excluded cross-section expected in the absence of a signal
(at 95\%\ confidence level).
  The variables $\left\{ x_{i}\right\} $ used to build the $\mathcal{L}_{R}$ functions 
 in the present analysis were \cite{DN189}: the visible energy (\Evis), visible mass (\Mvis), missing transverse momentum (\ptmiss),
 polar angle of the missing momentum, number of charged particles, total number of particles, 
acoplanarity, acollinearity, ratio of electromagnetic energy to total energy, percentage of total energy 
within 30$^\circ$ of the beam axis, 
kinematic information concerning the isolated photons, leptons and two most energetic charged particles and finally the jet characteristics. 

\subsection{Chargino analysis \label{sub:CHASEL}}

  The signal and background events were divided into four mutually exclusive topologies:
\begin{itemize}
\item{The \ll\ topology with no more than five charged particles and no isolated photons.}
\item{The \jjl\ topology with more than five charged particles and at least one isolated lepton and no isolated photons.}
\item{The \jjjj\ topology with more than five charged particles and no isolated photons or leptons.}
\item{The \rad\ topology with at least one isolated photon.}
\end{itemize}
The events in a given
topology are mostly events of the corresponding decay
channel, but events from other channels may also
contribute. For instance, for low mass difference, \DM, between the chargino and the lightest neutralino 
(and thus low visible energy) 
 some events with hadronic decays are selected in the leptonic topology, and some mixed decay  events with 
the isolated lepton unidentified enter into the hadronic topology. This migration effect tends to disappear 
as \DM\ increases. This effect was taken into account in the final efficiency and limit computations.

  The properties of the chargino decay products are mainly governed by the \DM\ value. For low \DM, 
the signal events are similar to \gamgam\ events, for high \DM\ to four-fermion final states
 (\WW, \ZZ,..) while for intermediate \DM\ values, the background is composed of many SM processes in comparable proportions.
 
  The signal events were simulated for 76 combinations of \XPM{1} and \XN{1} masses for five chargino mass values 
(\MXC{1}~$\approx$~94, 85, 70, 50  and 45~\GeVcc) and with \DM\ ranging from 3~\GeVcc\ to 70~\GeVcc. 
A total of 152000 chargino events (2000 per combination) was generated and passed through the complete simulation of 
the DELPHI detector. The kinematic properties (acoplanarity, \Evis, \ptmiss,..) of the signal events were studied 
in terms of their mean value and standard deviation, and six \DM\ regions were defined in order to have signal events with 
similar properties (table  \ref{tab:REGION}). 

\begin{figure}[ht]
\begin{center}
\begin{tabular}{cc}
\hspace*{-1.3cm}
\begin{minipage}[c]{8.0cm}
\epsfxsize=9.3cm  
\epsffile{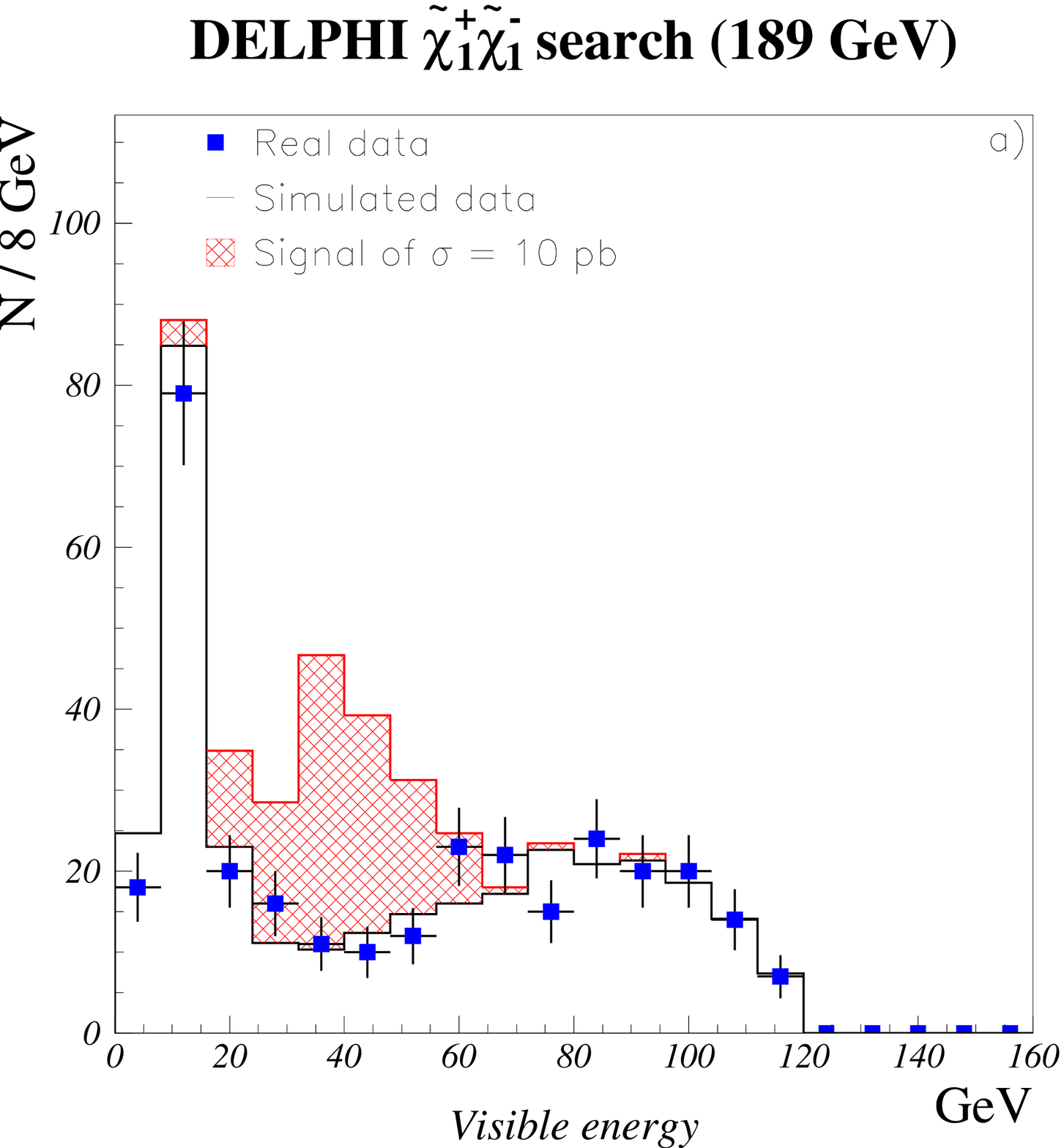}
\end{minipage}
&
\hspace*{0.4cm}
\begin{minipage}[c]{8.0cm}
\epsfxsize=9.3cm  
\epsffile{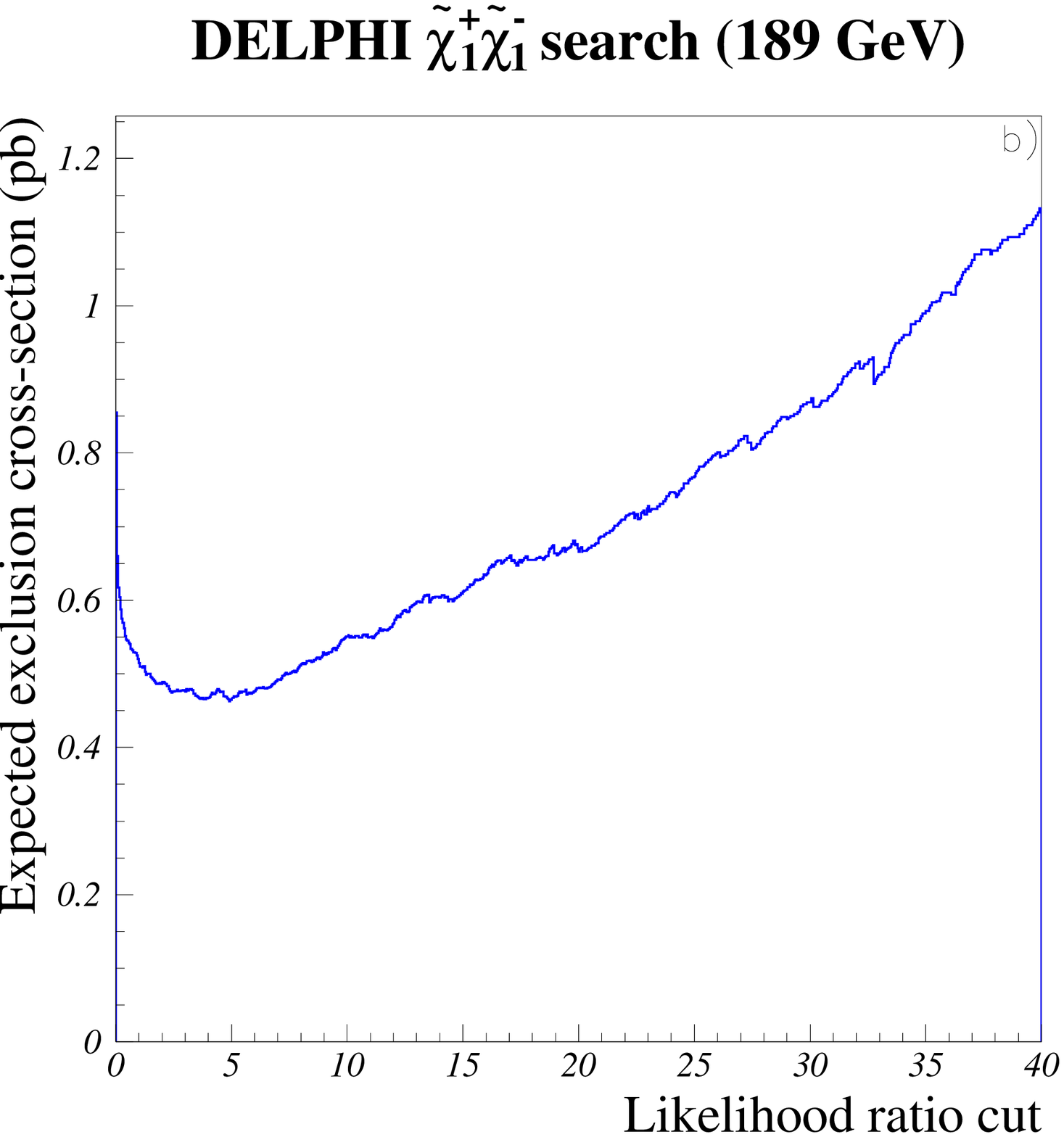}
\end{minipage}
\\
\end{tabular}
\vspace{0.5cm}\\
\begin{tabular}{cc}
\hspace*{-0.9cm}
\begin{minipage}[c]{8.0cm}
\hspace*{-0.5cm}
\epsfxsize=9.3cm  
\epsffile{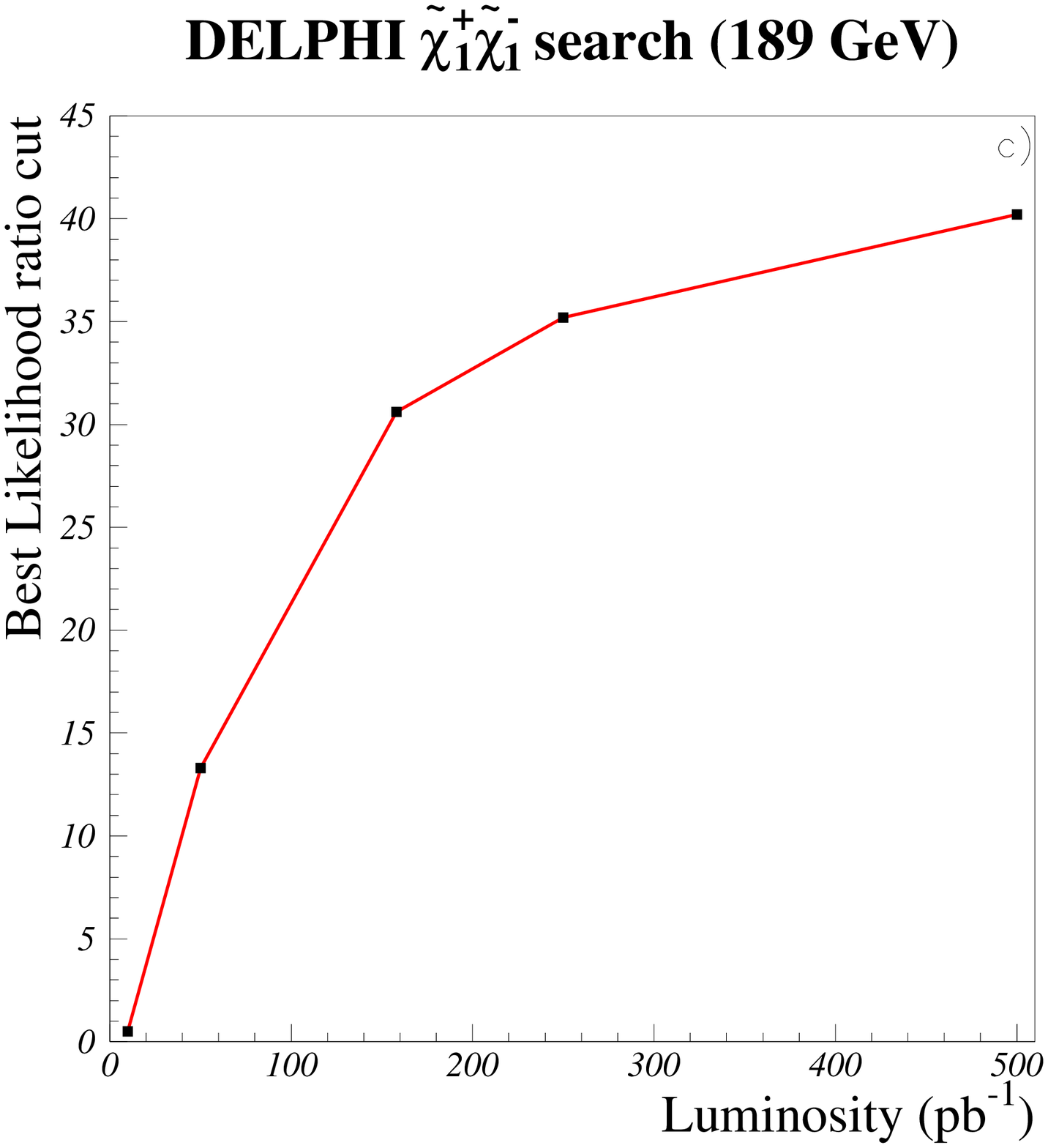}
\end{minipage}
&
\hspace*{0.01cm}
\begin{minipage}[c]{8.0cm}
\epsfxsize=9.3cm  
\epsffile{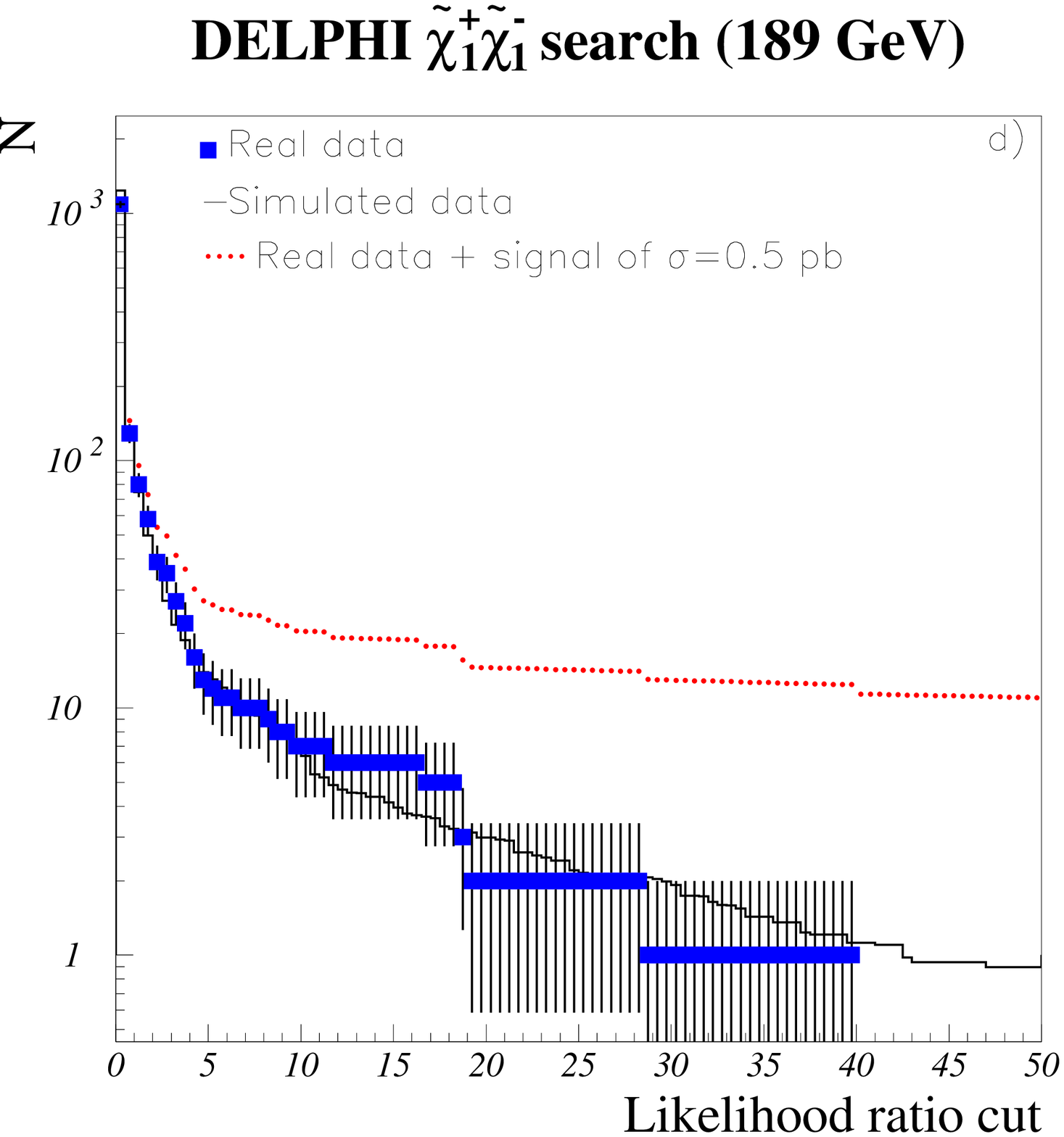}
\end{minipage}
\\
\end{tabular}
\vspace{0.5cm}
\caption{a) comparison between real data (squares), simulated background events (histogram) and a possible chargino signal,
\MXC{1}=94~\GeVcc\ \MXN{1}=40~\GeVcc, of 10 pb (hatched) at the preselection level and b) choice of
the best likelihood ratio cut for \DM\ $>$ 50~\GeVcc\  in the \ll\ topology. The dependence of the
optimum likelihood ratio cut as a function of the luminosity is shown in c) and the good agreement between
real (squares) and simulated (histogram) events as a function of the likelihood ratio cut is shown in d),
for 35 $\leq$ \DM\ $<$ 50~\GeVcc\ in the \jjl\ topology. A possible signal, \MXC{1}=94~\GeVcc\ \MXN{1}=54~\GeVcc,
of 0.5 pb added to real data is shown in figure d) by the dotted curve.} 
\label{fig:DATAMC}
\end{center}
\end{figure}

  In each of these 24 windows (four topologies, six \DM\ regions), a likelihood ratio function was defined.
The generation of these 24 functions was performed in five steps:

\begin{itemize}

\item{The signal distributions of all the variables used in this analysis (see section \ref{sub:LIKELI}) were built 
with signal events generated with parameter sets giving rise to charginos and neutralinos with masses in the 
corresponding \DM\ region. For each \DM\ region the events were classified according to the
above topological cuts. The background distributions were built with background events passing the same topological cuts.}
\item{Different preselection cuts, for each \DM\ region, were applied in order to reduce the high cross-section 
backgrounds (two-photon interactions and Bhabha events) and to generate the pdfs.
Fig.~\ref{fig:DATAMC}.a shows the distribution of the visible energy
for \DM\ $>$ 50~\GeVcc\  in the \ll\ topology for real and simulated events. 
The agreement is satisfactory, the normalization is absolute.
The pdfs were then generated as mentioned in~\ref{sub:LIKELI}.} 
\item{Then, to reduce statistical fluctuations a smoothing was performed 
by passing the 24 sets of pdfs for signal and background through a triangular filter~\cite{FILTER}.}
\item{In each window all the combinations of the pdfs were tested, starting from a minimal set of four variables. Every combination defined a $\mathcal{L}_{R}$ 
function (see section \ref{sub:LIKELI}) and 
a $\mathcal{L}_{R_{CUT}}$ computed in order to have the minimal expected excluded cross-section at 95\% C.L. (Fig.~\ref{fig:DATAMC}.b) using 
the monochannel Bayesian formula \cite{OBRA}. The parameters entering this computation were the 
 number of expected background events and the efficiency of the chargino selection. The efficiency of the chargino selection
 was defined in this case, as the number of events satisfying $\mathcal{L}_{R}\! >\! \mathcal{L}_{R_{CUT}}$ 
 divided by the 
total number of chargino events satisfying the topological cuts. Fig.~\ref{fig:DATAMC}.c shows the dependence of 
the optimum likelihood ratio cut on the integrated luminosity, 
 which demonstrates the importance of adjusting the cut to the luminosity.
Fig.~\ref{fig:DATAMC}.d shows the good agreement obtained
between real and simulated events as a function of the likelihood ratio cut,
for 35 $\leq$ \DM\ $<$ 50~\GeVcc\ in the \jjl\ topology.} 
\item{The combination of variables corresponding to the lowest excluded cross-section defined the $\mathcal{L}_{R}$ 
function and the $\mathcal{L}_{R_{CUT}}$ of this window.}
\end{itemize}
 Finally, the selection to be applied for SUSY
 models with \DM\ inside one such window was
 defined as a logical OR of the criteria for several
 windows, chosen to minimise the excluded  
 cross-section expected in the absence of a
 signal \cite{DN189}.

\begin{table}[ht]
 \begin{center}
   \begin{tabular}{|c|c|}
 \hline
\multicolumn{2}{|c|}{\DM\ regions} \\
 \hline
 1   &  3$\leq$\DM $<$ 5~\GeVcc \\
 \hline
 2   &  5$\leq$\DM $<$ 10~\GeVcc \\
 \hline
 3   &  10$\leq$\DM $<$ 25~\GeVcc \\
 \hline
 4   &   25$\leq$\DM $<$ 35~\GeVcc \\
 \hline
 5   &  35$\leq$\DM $<$ 50~\GeVcc \\
 \hline
 6   &   50~\GeVcc$\leq$\DM  \\
 \hline
\end{tabular}
\caption[.]{
\label{tab:REGION}
Definitions of the \DM\ regions.}
\end{center}
\end{table}

\section{Results \label{sec:RES}}

\subsection{Stable \XN{1}\ case \label{sec:RESSTABLE}}

\subsubsection{Efficiencies and selected events \label{sub:EFFSTABLE}}

  The total number of background events expected in the different topologies
is shown in table \ref{tab:EXPEV}, together with the
number of events selected in the data.

% ---------------------------------------------------------------------------
\begin{table}[hbtp]
\begin{center}
\begin{tabular}{|c||c|c|c|c||c|} 
\multicolumn{3}{l}{Stable \XN{1}} &
\multicolumn{3}{c}{}  \\
\hline
  &  &  &  &  & \\
{\large Topology:} & {\large \jjl} & {\large \ll} &  {\large \jjjj} &  {\large rad}  & {\large Total} \\ 
  &  &  &  &  & \\
\hline \hline
 & \multicolumn{5}{|c|}{\small{3 $\leq$ \DM\ $<$ 5~\GeVcc}  } \\ \hline
{\small Obs. events:}  & 0 & 46 & 1 & 4 & 51 \\
{\small Expect. events:} & 0.26 $^{+ 1.55 } _{- 0.07 }$ & 43.2  $^{+ 3.7  } _{- 2.4  }$ & 0.81 $^{+ 1.6  } _{- 0.13 }$ & 2.85 $^{+ 1.61 } _{- 0.32 }$ & 47.1  $^{+ 4.6  } _{- 2.4  }$ \\
\hline
 & \multicolumn{5}{|c|}{ \small{5 $\leq$ \DM\ $<$ 10~\GeVcc} } \\ \hline
{\small Obs. events:}  & 0 & 14 & 4 & 4 & 22 \\
{\small Expect. events:} & 0.26 $^{+ 1.55 } _{- 0.07 }$ & 14.3  $^{+ 2.5  } _{- 1.2  }$ & 2.27 $^{+ 1.81 } _{- 0.29 }$ & 2.85 $^{+ 1.61 } _{- 0.32 }$ & 19.7  $^{+ 3.8  } _{- 1.2  }$ \\
\hline
 & \multicolumn{5}{|c|}{ \small{10 $\leq$ \DM\ $<$ 25~\GeVcc }} \\ \hline
{\small Obs. events:}  & 0 & 25 & 9 & 4 & 38 \\
{\small Expect. events:} & 0.48 $^{+ 1.55 } _{- 0.08 }$ & 25.2  $^{+ 2.7  } _{- 1.4  }$ & 5.36 $^{+ 1.97 } _{- 0.49 }$ & 2.85 $^{+ 1.61 } _{- 0.32 }$ & 33.9  $^{+ 4.0  } _{- 1.6  }$ \\
\hline
 & \multicolumn{5}{|c|}{ \small{25 $\leq$ \DM\ $<$ 35~\GeVcc }} \\ \hline
{\small Obs. events:}  & 0 & 11 & 4 & 4 & 19 \\
{\small Expect. events:} & 0.26 $^{+ 1.52 } _{- 0.05 }$ & 13.0  $^{+ 2.0  } _{- 0.8  }$ & 5.81 $^{+ 1.59 } _{- 0.37 }$ & 2.85 $^{+ 1.61 } _{- 0.32 }$ & 21.9  $^{+ 3.4  } _{- 0.9  }$ \\
\hline
 & \multicolumn{5}{|c|}{ \small{35 $\leq$ \DM\ $<$ 50~\GeVcc }} \\ \hline
{\small Obs. events:}  & 0 & 24 & 21 & 2 & 47 \\
{\small Expect. events:} & 0.94 $^{+ 1.53 } _{- 0.14 }$ & 25.8  $^{+ 2.6  } _{- 1.1  }$ & 17.5  $^{+ 1.7  } _{- 0.7  }$ & 1.71 $^{+ 1.56 } _{- 0.19 }$ & 45.9  $^{+ 3.6  } _{- 1.4  }$ \\
\hline
 & \multicolumn{5}{|c|}{ \small{50~\GeVcc\ $\leq$ \DM }} \\ \hline
{\small Obs. events:}  & 1 & 24 & 27 & 2 & 54 \\
{\small Expect. events:} & 1.32 $^{+ 1.56 } _{- 0.17 }$ & 19.7  $^{+ 2.0  } _{- 0.9  }$ & 21.8  $^{+ 1.9  } _{- 0.8  }$ & 1.71 $^{+ 1.56 } _{- 0.19 }$ & 44.8  $^{+ 3.5  } _{- 1.2  }$ \\
\hline
 & \multicolumn{5}{|c|}{ \small{TOTAL (logical OR between different \DM\ windows)}} \\ \hline
{\small Obs. events:}  & 1 & 70 & 36 & 5 & 112 \\
{\small Expect. events:} & 1.32 $^{+ 1.56 } _{- 0.17 }$ & 66.7  $^{+ 3.9  } _{- 2.7  }$ & 25.6  $^{+ 2.2  } _{- 1.0  }$ & 3.57 $^{+ 1.63 } _{- 0.37 }$ & 97.2  $^{+ 5.0  } _{- 2.9  }$ \\
\hline
\end{tabular}
\end{center}
\caption[.]{
\label{tab:EXPEV}
The number of events observed in data and the expected number of
background events in the different chargino search topologies under
the hypothesis of a stable \XN{1}\ (section \ref{sub:CHASEL}).}
\end{table}
% ---------------------------------------------------------------------------

\begin{figure}[ht]
\begin{center}
\vspace*{0.0cm}
\begin{tabular}{c}
\epsfxsize=17.0cm
\epsffile{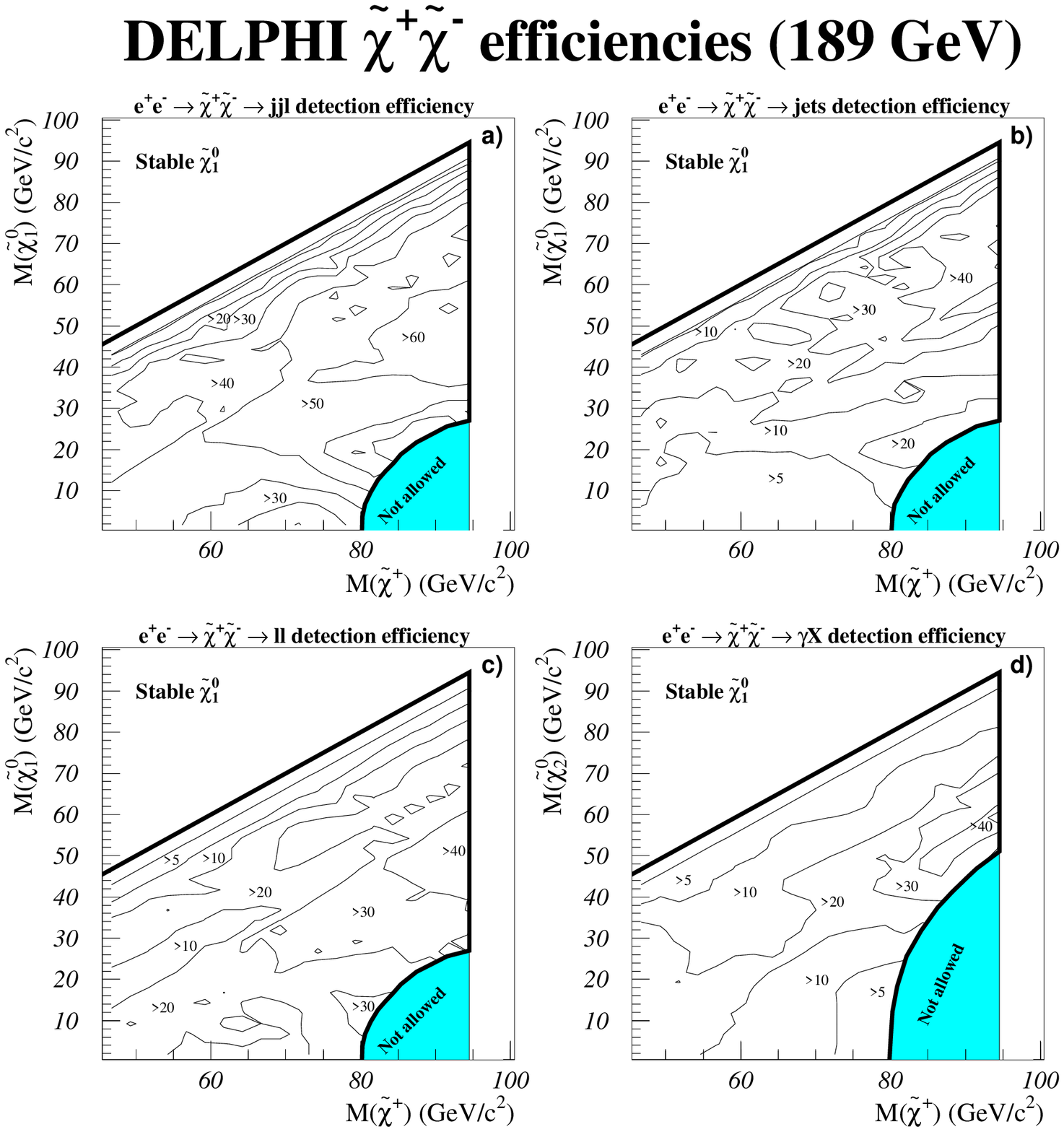}
\end{tabular}
\caption{Chargino pair production detection efficiencies~(\%) for the four decay channels a) \jjl,
b) \jjjj, c) \ll\ and d) \rad, at 189~GeV in the ($\MXc,\MXn$) plane. A
stable \XN{1}\ is assumed. The shaded areas are disallowed in the MSSM scheme.}
\label{fig:CHAEFF}
\end{center}
\end{figure}

  The efficiencies of the chargino selection in the four topologies were computed separately for the 76 MSSM points 
using the $\mathcal{L}_{R}$ function and the $\mathcal{L}_{R_{CUT}}$ of the corresponding topology and \DM\ region.
  To pass from the efficiencies of the chargino selection in the four topologies to the efficiencies in the four 
decay channels, all the migration effects were computed for all the generated points of the signal simulation. 
Then the efficiencies of the selection in the four decay channels were interpolated 
in the (\MXC{1},\MXN{1}) plane using the same method as in Ref. \cite{PAP172}. 
When the interpolation was not possible (for $\MXC{1} \sim 80~\GeVcc$ and $\MXN{1} \sim 0~\GeVcc$) 
an extrapolation was used. 
These efficiencies as functions of \MXC{1}\ and \MXN{1} are shown in Fig.~\ref{fig:CHAEFF}.
 
  All the selected events in the real data are compatible with the expectation from the 
background simulation. As
no evidence for a signal is found, exclusion limits are set at 95\% C.L.
using the multichannel Bayesian formula \cite{OBRA} taking into account the branching ratio and 
the efficiency of each decay channel.

\subsubsection{Limits \label{ssub:LIMSTABLE}}

\underline{\em Limits on chargino production}\nopagebreak[4]
\vskip 2mm

  The simulated points were used to parametrize the
efficiencies of the chargino selection criteria described in section
\ref{sub:CHASEL} in terms of \DM\ and the mass of the
chargino (see section \ref{sub:EFFSTABLE}). Then a large number of SUSY points were investigated and the
values of \DM, the chargino and neutralino masses and the various decay
branching ratios were determined for each point.
By applying the appropriate efficiency (from the interpolation) and branching ratios and cross-sections for
each channel decay (computed by {\tt SUSYGEN}), the number of expected signal events can be calculated. Taking into account the expected background and the number of observed 
events, the corresponding 
point in the MSSM parameter space ($\mu $, $M_2$, \tanb) can be excluded if the number of 
expected signal events is greater than the upper limit at 95\% C.L. on the number of observed events 
of the corresponding \DM\ region.

\begin{figure}[ht]
\begin{center}
\mbox{\epsfysize=512.15pt\epsffile{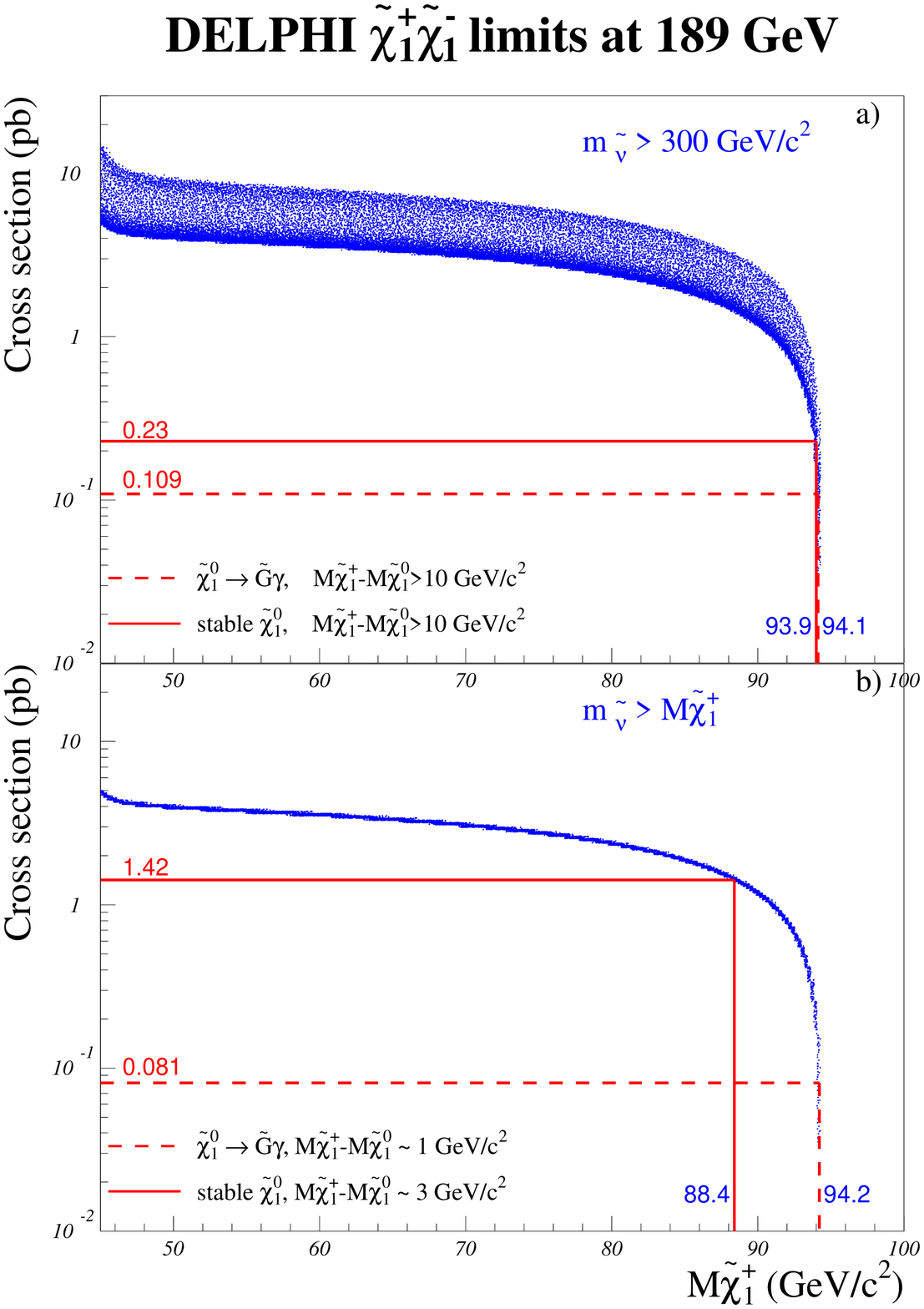}}
\vspace{14.23pt}
\caption[]{Expected cross-sections in pb at 189 GeV (dots)
versus the chargino mass in a) 
in the non-degenerate case (\DM $>$ 10~\GeVcc) and b) the degenerate case  
(\DM $\sim$ 3~\GeVcc). The spread in the dots originates from the random scan over the
parameters $\mu$ and M$_2$.
A heavy sneutrino (\msneu $>$ 300~\GeVcc) has been assumed in a) and $\msneu\! >\!\MXC{1}$ in b).
The minimal cross-sections below the mass limits are indicated by the horizontal lines.}
\label{fig:SMCHAL}
\end{center}
\end{figure}

  Fig.~\ref{fig:SMCHAL} shows
the chargino production cross-sections as obtained in the MSSM
at \rs~=~189~\GeV\ for different chargino masses for the non-degenerate (\DM\ $>$ 10~\GeVcc)
and degenerate cases  (\DM~=~3~\GeVcc) .
The parameters $M_2$ and $\mu $ were varied randomly in the ranges
0~\GeVcc\ $< M_2 <$ 3000~\GeVcc\ and --200~\GeVcc\ $ < \mu < $ 200~\GeVcc\
for three fixed different values of \tanb,
namely 1, 1.5 and 35. The random generation of the parameters led to 
an accuracy on the mass limit computation of the order
of 10~\MeVcc. Two different cases were considered for the sneutrino mass:
$\msnu\! >\! 300~\GeVcc$ (in the non-degenerate case) and $\msnu\!>\!\MXC{1}$  
(in the degenerate case).

  To derive the chargino mass limits, constraints on the process
$\Zn \to \XN{1}\XN{2} \to \XN{1}\XN{1}\gamma$ were also included.
These were derived from the DELPHI results on single-photon production 
at LEP~1 \cite{SINGLEGAMMA1}.

\begin{figure}[ht]
\begin{center}
\mbox{\epsfysize=16.0cm\epsffile{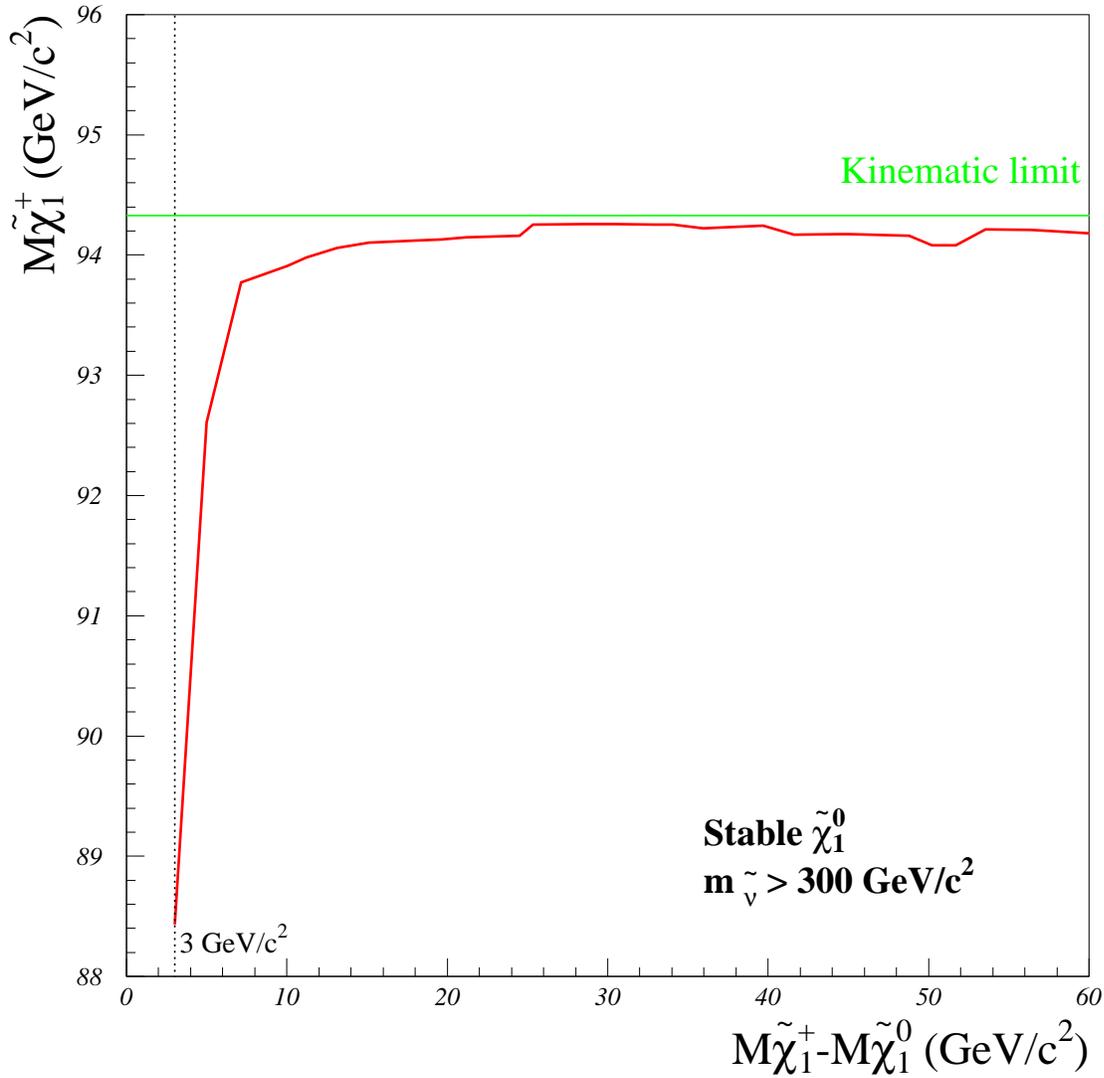}}
\vspace{0.5cm}
\caption[]{
The chargino mass limit as function of the \DM\ value
under the assumption of a heavy sneutrino.
The limit applies to the
case of a stable \XN{1}.
The straight horizontal line shows the kinematic limit.
}
\label{fig:MCHADM}
\end{center}
\end{figure}

\begin{figure}[ht]
\begin{center}
\mbox{\epsfysize=16.0cm\epsffile{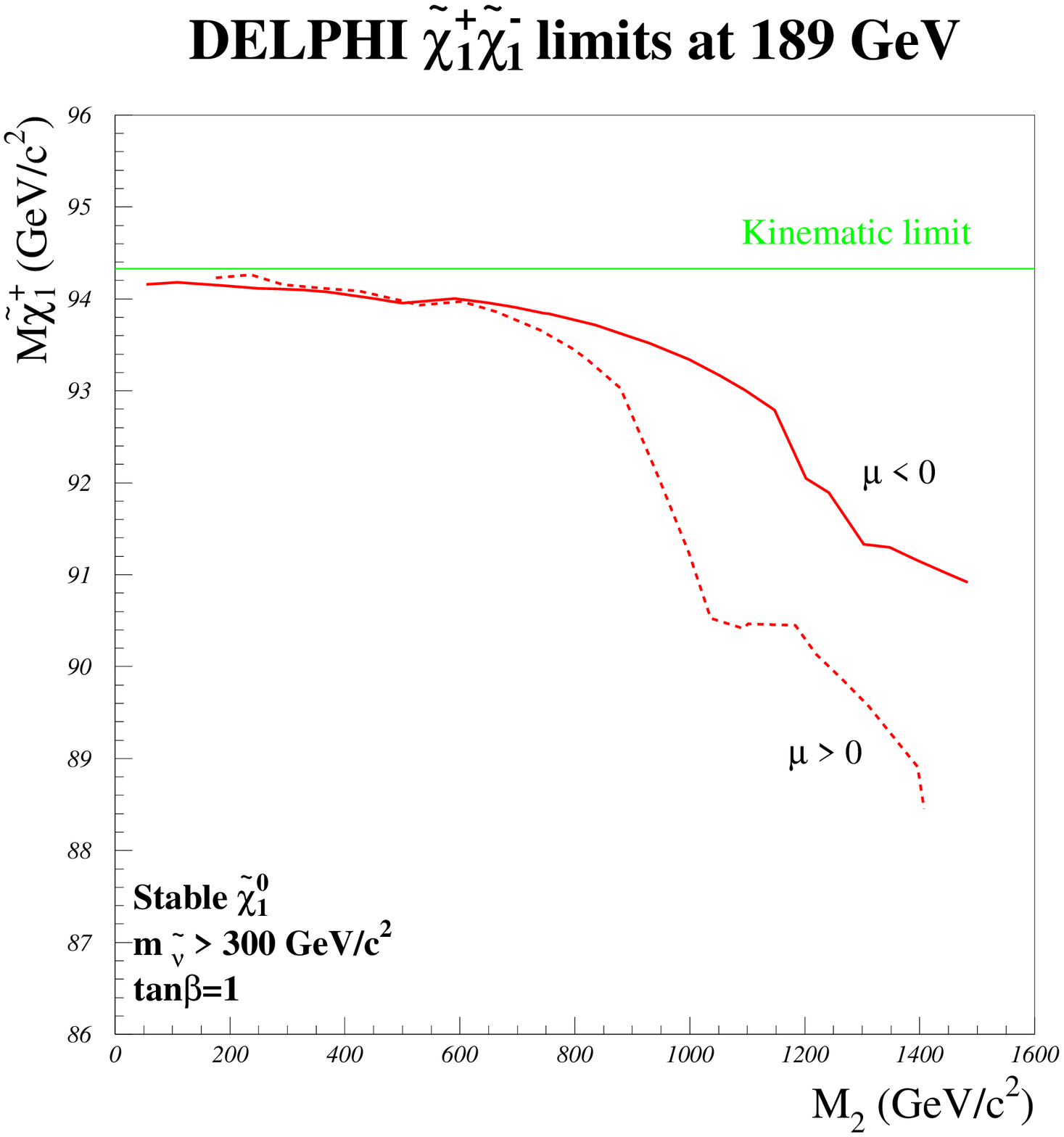}} 
\vspace{0.5cm}
\caption[]{
The chargino mass limit as function of $M_2$ for \tanb\ = 1,
under the assumption of a heavy sneutrino (\msneu $>$ 300~\GeVcc).
%KH for small values of $M_2$.
The straight horizontal line shows the kinematic limit in the production.
The limit applies in the case of a stable \XN{1}.
}
\label{fig:MCHAM2}
\end{center}
\end{figure}

The chargino mass limits are summarized in Table~\ref{tab:CHALIM}.
The table also gives, for each case, the minimal MSSM cross-section
for which \MXC{1} is below the corresponding mass limit.
These cross-section values are also displayed in Fig.~\ref{fig:SMCHAL}. The chargino mass limits
versus \DM\ and versus $M_2$, assuming a heavy sneutrino, are shown in Figs.~\ref{fig:MCHADM}
and \ref{fig:MCHAM2}, respectively. The behaviour of the curve in Fig.~\ref{fig:MCHADM} depends 
very weakly on the relation between $M_1$ and $M_2$. 
Note that in Fig.~\ref{fig:MCHAM2}, for a fixed high value of $M_2$, the 
chargino mass limit is lower for positive $\mu$ than for negative $\mu$. This is due to higher
degeneracy for positive $\mu$ than for negative $\mu$, for a fixed value of $M_2$.

  In the non-degenerate case (\DM\ $>$ 10~\GeVcc) with a large sneutrino mass ($>$~300~\GeVcc),
the lower limit for the chargino ranges between 93.9~\GeVcc\ (for a mostly higgsino-like
chargino) and 94.2~\GeVcc\ (for a mostly wino-like chargino). The minimal excluded MSSM 
cross-section at \rs~=~189~\GeV\ is 0.23~pb, deriving from a chargino mass limit of 93.9~\GeVcc. 
For \DM\ $>$ 20~\GeVcc, the lower limit for the chargino mass ranges between 94.1~\GeVcc\ and 
94.2~\GeVcc. In this case the minimum excluded MSSM cross-section
at \rs~=~189~\GeV\ is 0.13~pb.

  In the degenerate case (\DM~=~3~\GeVcc), the cross-section does not depend
significantly on the sneutrino mass, since
the chargino is higgsino-like under the assumption of
gaugino mass unification.
The lower limit for the chargino mass, shown in Fig.~\ref{fig:SMCHAL},
is 88.4~\GeVcc. The minimal excluded cross-section is in this case 1.42~pb.

  The systematic error on the given mass limits is less than 0.5\% for \DM~=~3~\GeVcc\ and
 less than 0.1\% for \DM~$>$~20~\GeVcc.

%-------------------------------------------------------------------

\begin{table}[ht]
\begin{center}
%\vspace*{0.4in}
\begin{tabular}{||c|c||c|c|c||} \hline \hline
           &          &                      &                &                  \\
{\bf Case} & $\msneu$ & $M^{min}_{\chi^\pm }$ & $\sigma^{max}$ & ${\rm N}_{95\%}$ \\
           &          &                      &                &                  \\
           & (\GeVcc) & (\GeVcc)             & (pb)           &                  \\ \hline \hline
\multicolumn{5}{||c||}{ } \\
\multicolumn{5}{||c||}{Stable \XN{1}} \\
\multicolumn{5}{||c||}{ } \\\hline 
                  &         &      &      &                \\
$\DM > 20$~\GeVcc & $>$~300 & 94.1 & 0.13 & 10.6 \\
                  &         &      &      &       \\
$\DM > 10$~\GeVcc & $>$~300 & 93.9 & 0.23 & 13.2 \\
                  &         &      &      &       \\
$\DM = 3$~\GeVcc  & $>$~\MXC{1} & 88.4 & 1.42 & 12.3  \\
                  &         &      &      &                \\ \hline 
\multicolumn{5}{||c||}{ } \\
\multicolumn{5}{||c||}{Unstable \XN{1}} \\
\multicolumn{5}{||c||}{ } \\\hline 
                  &         &      &      &                \\
$\DM > 10$~\GeVcc & $>$~300   & 94.1 & 0.11 & 8.6  \\
                  &         &      &      &                \\
$\DM = 1$~\GeVcc  & $>$~\MXC{1}    & 94.2 & 0.08  & 7.1 \\
                  &         &      &      &                \\ \hline \hline 
\end{tabular}
\end{center}
\vspace{0.5 cm}
\caption{95\% confidence level lower limits for the chargino mass, the corresponding
pair production cross-sections at 189~GeV and the 95\% confidence 
level upper limit on number of observed events, for the non-degenerate and a highly 
degenerate cases. The scenarios of a stable \XN{1}\ and \XN{1} $\to$ \Gino$\gamma$ are considered.}
\label{tab:CHALIM}
\end{table}
% ---------------------------------------------------------------------------
%-------------------------------------------------------------------

\vskip 2mm
\underline{\em Limits on MSSM parameters and neutralino mass}\nopagebreak[4]
\vskip 2mm

\begin{figure}[htbp]
\begin{center}
%\vspace*{-0.5cm}
\begin{tabular}{c}
\epsfxsize=15.0cm
\epsffile{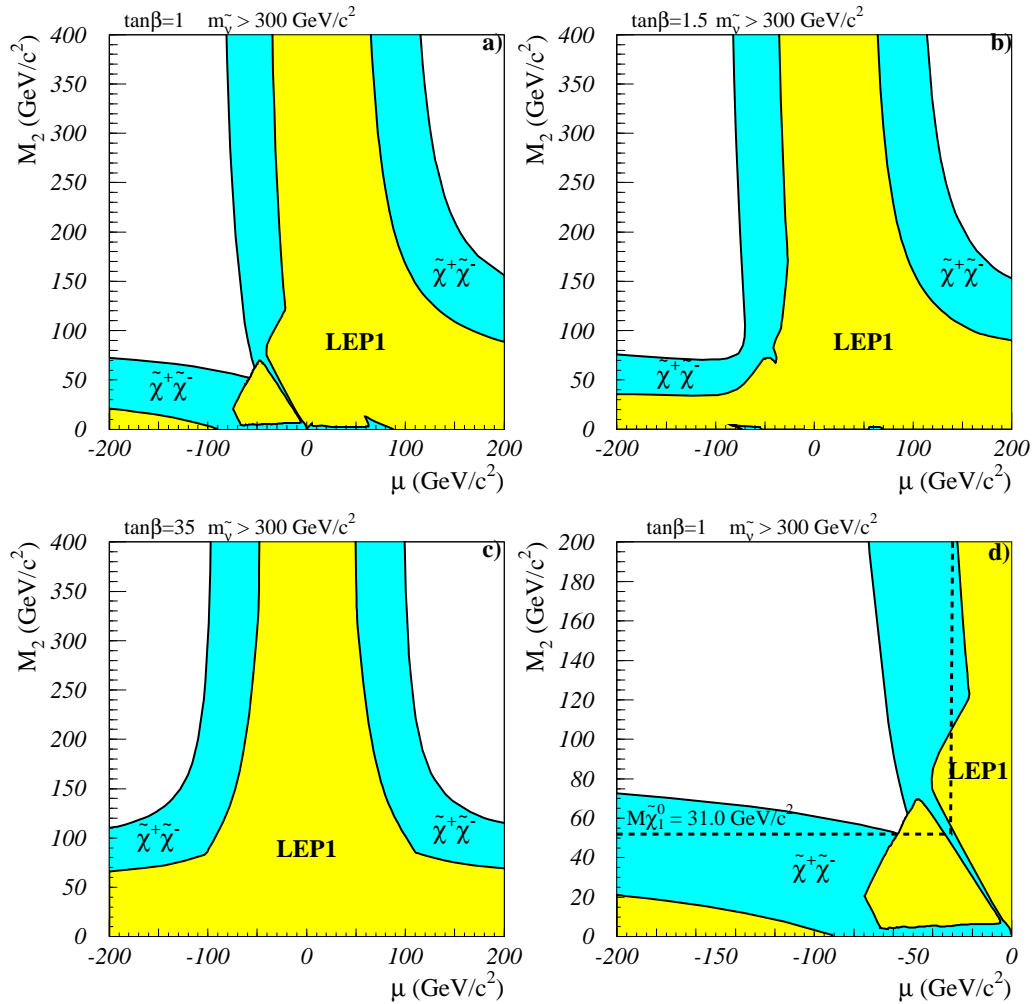}
\end{tabular}
\vspace*{0.5cm}
\caption[]{a), b), and c), regions excluded at 95~\% confidence level in the ($\mu$, M$_2$) plane at $\sqrt{s}$ = 189 GeV under the assumption of a
heavy sneutrino for \tanb~=~1, 1.5 and 35. The dark shading shows the region excluded by the chargino search 
and the light shaded region is the one excluded by LEP1. The constant mass curve for the LSP mass limit is shown in d) by the dashed line, for \tanb~=~1.}
\label{fig:neulim_m2m}
\end{center}
\end{figure}

The exclusion regions in the ($\mu,M_2$) plane for 
\tanb~=~1, 1.5 and 35  are shown in Fig.~\ref{fig:neulim_m2m}.a, \ref{fig:neulim_m2m}.b, and \ref{fig:neulim_m2m}.c,
assuming a heavy sneutrino (\msneu $>$ 300~\GeVcc).
These limits, based on data taken at \rs~=189~GeV,
improve on previous limits at lower energies, and represent a significant
increase in range as compared to LEP~1 results \cite{LEP1LIM}. 

DELPHI limits on the process $\Zn \to \XN{1}\XN{2} \to \XN{1}\XN{1}\gamma$ 
were derived from the single-photon
search at LEP~1~\cite{SINGLEGAMMA1}. These limits marginally extends part of the region covered by
the chargino search at low \tanb\ for small $M_2$ and negative $\mu$ (Fig.~\ref{fig:neulim_m2m}.d). The 
exclusion region obtained depends strongly on the the assumed GUT relation between
$M_1$ and $M_2$. 

\begin{figure}[ht]
\begin{center}
\mbox{\epsfysize=455.25pt\epsffile{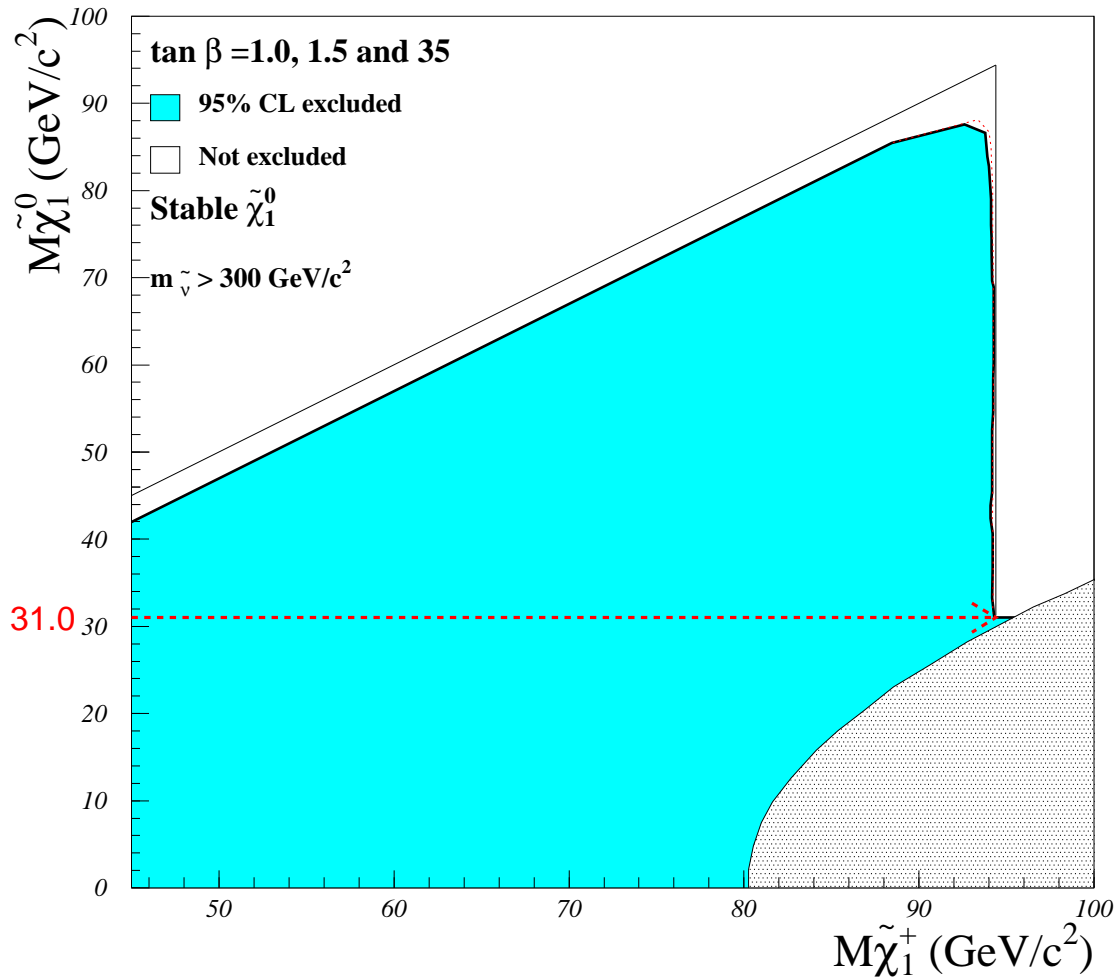}}
\vspace{14.23pt}
\caption[]{
Region excluded at 95\% confidence level in the plane of the mass
of the lightest neutralino versus that of the lightest chargino
under the assumption of a heavy sneutrino, for \tanb~=~1.0, 1.5 and 35.
The thin lines show the kinematic limits in the production and
the decay. The dotted line (partly hidden by the shading) shows 
the expected exclusion limit. The lightly shaded region is not
allowed in the MSSM. The limit applies in the
case of a stable \XN{1}. The mass limit on the lightest neutralino is 
indicated by the horizontal dashed line.
The excluded region outside the kinematic limit is obtained from the limit 
on \XN{1}\XN{2}\ production at the Z resonance derived from the single-photon search.
}
\label{fig:NEUCHA}
\end{center}
\end{figure}

The exclusion regions in the ($\mu$,$M_2$) plane
can be translated into a limit on the mass of the lightest neutralino also 
shown in the (\MXC{1},\MXN{1})-plane in Fig.~\ref{fig:NEUCHA}.
A lower limit of 31.0~\GeVcc\ on the lighest neutralino mass is obtained, valid for
\tanb\ $\ge$ 1 and a heavy sneutrino. This limit is reached for
\tanb~=~1, $\mu~=~-57.8$~\GeVcc, M$_2$~=~52.05~\GeVcc. 
%KH Using the limits on chargino production and on LEP 1 single-photon production the
%KH excluded region in the plane of neutralino mass versus chargino mass was
%KH determined, as shown in Fig.~\ref{fig:NEUCHA}. A heavy sneutrino was assumed. 
The small excluded
region outside the chargino kinematic limit in Fig.~\ref{fig:NEUCHA} derives from the single-photon
search at LEP~1~\cite{SINGLEGAMMA1}. In the same figure, the vertical dotted line (partly hidden by the shading) shows 
the expected exclusion limit.

\subsection{Unstable \XN{1}\ case \label{sub:RESUNSTAB}}

\subsubsection{Efficiencies and selected events \label{ssub:EFFUNSTAB}}

\begin{figure}[ht]
\begin{center}
\vspace*{0.0cm}
\begin{tabular}{c}
\epsfxsize=16.0cm
\epsffile{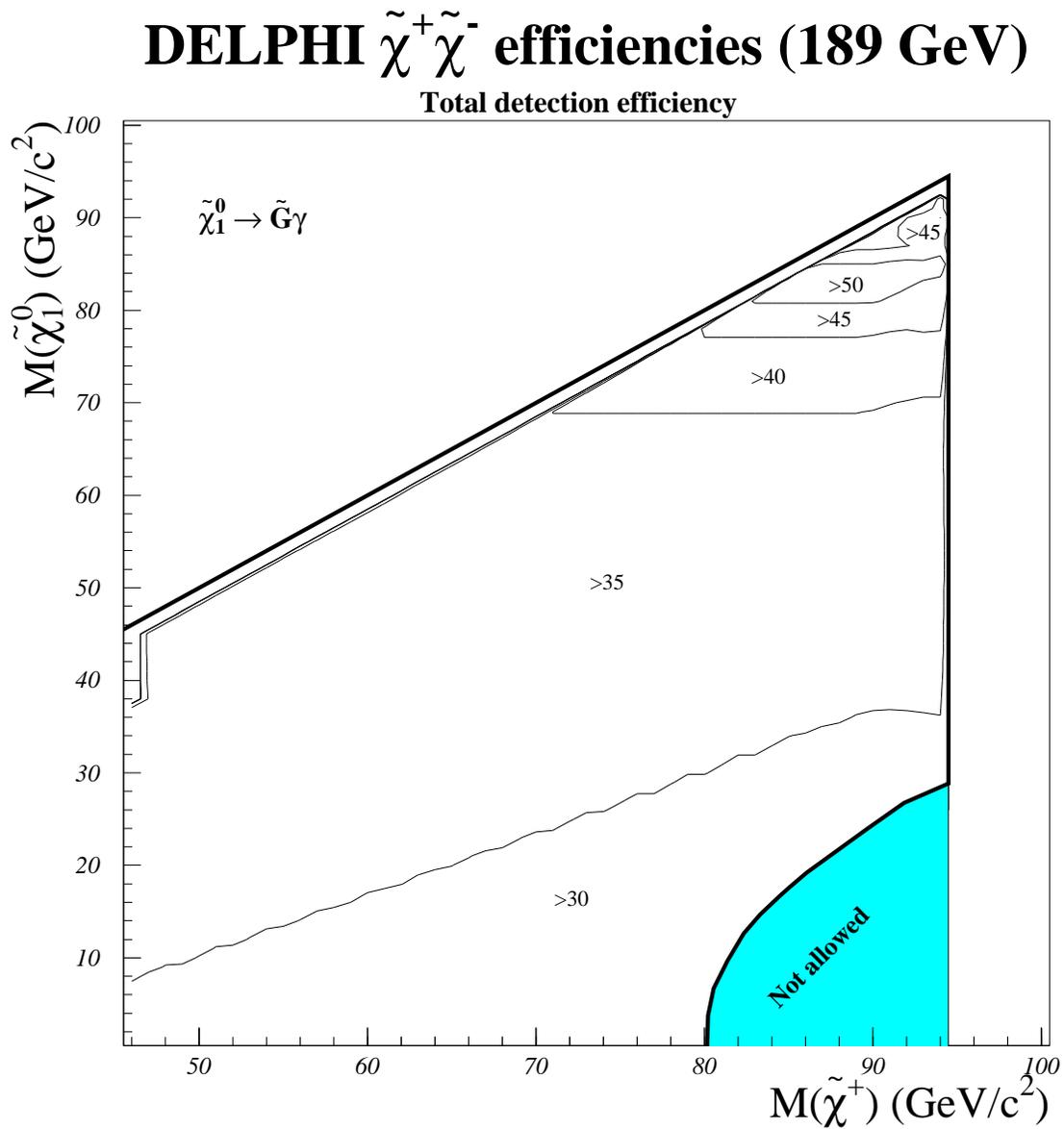}
\end{tabular}
\caption{Chargino pair production detection efficiency~(\%) at 189~GeV in the ($\MXc,\MXn$) plane.
An unstable \XN{1}\ is assumed. The shaded areas are disallowed by the MSSM scheme.}
\label{fig:CHAEFFGAM}
\end{center}
\end{figure}

  The efficiency of the chargino selection for an unstable \XN{1}\ 
decaying into a photon and a gravitino
was calculated from a total of 78000 events generated using the same combinations of
\MXC{1} and \MXN{1} as in the stable \XN{1}\ scenario.
As mentioned in \cite{PAP172}, the same selection applies to all
topologies.
  The efficiency, as shown in Fig.~\ref{fig:CHAEFFGAM}, varies only weakly
with \DM\ so only three \DM\ windows were used in this case.
Note that, due to the presence of the photons from the neutralino decay,
the region of high degeneracy (down to \DM~=~1~\GeVcc) is fully covered.

  The total number of background events expected in the three different \DM\ ranges
is shown in table \ref{tab:EXPEVGRA}, together with the
number of events selected in the data. 24 events were found in the data, with a
total expected background of 19.9 $\pm $1.9. 
Since no evidence for a signal was found, exclusion limits were set.

\subsubsection{Limits \label{ssub:LIMUNSTAB}}

  The chargino cross-section limits corresponding to the case where
the neutralino is unstable and decays via
$\XN{1} \rightarrow \Gino \gamma$
were computed as explained in section \ref{ssub:LIMSTABLE} and
are shown in Fig.~\ref{fig:SMCHAL} and in table \ref{tab:CHALIM}.
  In the non-degenerate case the chargino mass limit at 95\% C.L.
is 94.1~\GeVcc\ for a heavy sneutrino, 
while in the ultra-degenerate case (\DM~=~1~\GeVcc) the limit is 94.2~\GeVcc.
The minimal MSSM cross-sections excluded by the above mass limits are 0.109~pb in the 
non-degenerate case and 0.081~pb in the ultra-degenerate case.

% ---------------------------------------------------------------------------
\begin{table}[ht]
\begin{center}
\begin{tabular}{|c||c|c|c|}
\multicolumn{2}{l}{Unstable \XN{1} } &
\multicolumn{2}{c}{} \\
\hline
 & & & \\
   & \DM\ $>$ 10 \GeVcc\ & 5 $\le$ \DM\ $\le$ 10 \GeVcc\ & \DM\ $<$ 5 \GeVcc\ \\ 
 & & & \\
\hline \hline
Obs. events:      &    14              &  6                 &  4              \\
Expect. events:  &  15.1 $^{+ 1.8} _{- 0.8}$ & 2.6 $^{+ 1.77} _{- 0.32}$ & 2.2 $^{+ 1.73} _{- 0.26}$ \\ 
\hline
\end{tabular}
\end{center}
\caption[.]{
\label{tab:EXPEVGRA}
The number of events observed and the expected number of
background events in the different \DM\ cases under
the hypothesis of an unstable \XN{1}\ (section \ref{sub:CHASEL}).}
\end{table}
% ---------------------------------------------------------------------------

\section{Summary \label{sec:SUMMARY}}
Searches for charginos at \rs~=~189~GeV allow the
exclusion of a large domain of SUSY parameters, cross-sections, and masses, at
95\% confidence level.

  Assuming a
difference in mass between chargino and neutralino, \DM,  of 10~\GeVcc\ or
more, and a sneutrino heavier than 300~\GeVcc, the existence of a chargino
lighter than 93.9~\GeVcc\ can be excluded. If a gaugino-dominated
chargino is assumed in addition, the kinematic limit is reached.
If \DM\ is 3~\GeVcc, the lower limit on the
chargino mass becomes 88.4~\GeVcc, assuming a sneutrino heavier than the chargino.

A lower limit of 31.0~\GeVcc\ on the lightest neutralino mass is obtained assuming a
heavy sneutrino and $M_1/M_2 \approx 0.5$,
using the obtained chargino exclusion regions and including DELPHI results~\cite{SINGLEGAMMA1}
on the process $\Zn \to \XN{1}\XN{2} \to \XN{1}\XN{1}\gamma$~\ . 

  A specific \XP{1}\XM{1} production search was performed assuming the decay of the
lightest neutralino into a photon and a gravitino, giving
somewhat more stringent limits on cross-sections and masses than in the case of
a stable \XN{1}: \MXC{1}~$>$~94.1~\GeVcc\ for large \DM\ and \MXC{1}~$>$~94.2~\GeVcc\ for $\DM = 1$~\GeVcc.

\newpage
%         Modified on 04-06-1999 by dimartino
%-------------------------------------------------------------------
\subsection*{Acknowledgements}
\vskip 3 mm
 We are greatly indebted to our technical 
collaborators, to the members of the CERN-SL Division for the excellent 
performance of the LEP collider, and to the funding agencies for their
support in building and operating the DELPHI detector.\\
We acknowledge in particular the support of \\
Austrian Federal Ministry of Science and Traffics, GZ 616.364/2-III/2a/98, \\
FNRS--FWO, Belgium,  \\
FINEP, CNPq, CAPES, FUJB and FAPERJ, Brazil, \\
Czech Ministry of Industry and Trade, GA CR 202/96/0450 and GA AVCR A1010521,\\
Danish Natural Research Council, \\
Commission of the European Communities (DG XII), \\
Direction des Sciences de la Mati$\grave{\mbox{\rm e}}$re, CEA, France, \\
Bundesministerium f$\ddot{\mbox{\rm u}}$r Bildung, Wissenschaft, Forschung 
und Technologie, Germany,\\
General Secretariat for Research and Technology, Greece, \\
National Science Foundation (NWO) and Foundation for Research on Matter (FOM),
The Netherlands, \\
Norwegian Research Council,  \\
State Committee for Scientific Research, Poland, 2P03B06015, 2P03B1116 and
SPUB/P03/178/98, \\
JNICT--Junta Nacional de Investiga\c{c}\~{a}o Cient\'{\i}fica 
e Tecnol$\acute{\mbox{\rm o}}$gica, Portugal, \\
Vedecka grantova agentura MS SR, Slovakia, Nr. 95/5195/134, \\
Ministry of Science and Technology of the Republic of Slovenia, \\
CICYT, Spain, AEN96--1661 and AEN96-1681,  \\
The Swedish Natural Science Research Council,      \\
Particle Physics and Astronomy Research Council, UK, \\
Department of Energy, USA, DE--FG02--94ER40817. \\
%=========================================================================%
%=========================================================================%

% --------------------------------------------------------------------

\newpage

\newpage

\newpage

\newpage

\newpage

\newpage

\newpage

\newpage

% --------------------------------------------------------------------

\end{document}